# Morphology and Dynamics of Self-interstitial Clusters in Irradiated Nickel

Ajay Annamareddy[1,*], Ibrahim Momohjimoh[1], Hangyu Li[2], Kevin G. Field[2], Paul M. Voyles[1], Dane Morgan[1,*]

[1]*Department of Materials Science and Engineering, University of Wisconsin-Madison, Madison, WI 53706, USA*

[2]*Department of Nuclear Engineering and Radiological Sciences, University of Michigan, Ann Arbor, MI 48109, USA*

[*]Corresponding authors: ajeykrishna@gmail.com and ddmorgan@wisc.edu

Self-interstitial atom (SIA) clustering is a key early step in radiation damage evolution in face-centered cubic (FCC) metals, governing defect transport, recombination, and the long-term microstructural response of irradiated alloys. We combine molecular dynamics (MD) simulations and high-speed (>1000 frames/s) *in situ* transmission electron microscopy (TEM) to investigate the structure, energetics, and migration dynamics of SIA clusters in FCC Ni. MD simulations show that interstitials initially form disordered dumbbell clusters that evolve into either sessile Frank loops or glissile perfect (prismatic) loops; the latter progressively reorganize into compact ordered configurations with increasing mobility. Direct construction of both loop types over a wide size range, validated against MD-relaxed structures, shows that perfect loops are thermodynamically favored over Frank loops for cluster sizes $N \geq 14$, where $N$ is the number of SIAs, with the energetic advantage increasing with cluster size. Nevertheless, substantial kinetic barriers allow Frank loops to persist as metastable defects. For perfect loops, diffusion coefficients computed over $N$ = 16–400 reveal a nearly size-independent migration barrier of ~0.02 eV, while the diffusion prefactor decreases approximately as $N^{-0.54}$. Trajectory analysis reveals a non-rigid, row-wise relay mechanism in which the number of participating atoms increases systematically with loop size, accounting for much of the observed prefactor scaling. Sub-millisecond *in situ* TEM observations reveal intermittent loop motion at velocities higher than previously observed but still several orders of magnitude below the intrinsic mobilities predicted by MD, indicating migration through a heterogeneous energy landscape of mobile and pinned states.

## 1. Introduction

Face-centered cubic (FCC) metals and alloys, particularly Ni-based alloys and austenitic steels, play critical roles in nuclear energy systems due to their combination of high-temperature strength, corrosion resistance, and microstructural stability, and are routinely used in steam generators, heat exchangers, and other structural reactor components operating under demanding conditions [1–3]. Energetic particle irradiation in these materials can generate displacement cascades that produce both isolated defects and clusters of vacancies and self-interstitial atoms (SIAs) [4–7]. The subsequent evolution of these vacancy and interstitial clusters is strongly asymmetric. Vacancy clusters are generally immobile at typical operating temperatures, whereas interstitial clusters can reorganize into glissile perfect (prismatic) dislocation loops with Burgers vector $\boldsymbol{b} = 1/2\langle 110 \rangle$, composed of crowdions that migrate one-dimensionally along the crowdion axis. This fast interstitial transport strongly influences the spatial separation, recombination, and accumulation of irradiation-induced defects and plays an important role in the development of irradiation-induced dislocation networks [6,8]. Interstitial cluster mobility is a key, although poorly constrained, input to models of microstructural evolution in irradiated alloys. Consequently, understanding the dynamics of interstitial clusters is essential for connecting atomistic defect production to the long-term radiation response of FCC materials.

In addition to glissile perfect loops, irradiation-induced interstitial clusters can also form sessile Frank loops, characterized by $\boldsymbol{b} = 1/3\langle 111 \rangle$ and a faulted $\{111\}$ habit plane. Frank loops can transform into perfect loops through an unfaulting process, and the relative stability of these two defect structures has been the subject of extensive experimental and computational study [9–12]. Previous efforts to identify a well-defined crossover size between Frank and perfect loops have yielded conflicting conclusions [10], but the energetics of this transformation remain important for understanding the defect populations that develop under irradiation. In this work, we compare the energetics of Frank and perfect loops in FCC Ni over a broad range of cluster sizes.

The mobility of glissile interstitial loops is particularly important because it governs defect transport, loop interactions, and the formation of extended defect structures during irradiation. Atomistic simulations and experiments have shown that glissile interstitial loops in FCC metals can migrate one-dimensionally with exceptionally low activation barriers, yielding diffusion coefficients far larger than those of most other irradiation-induced defects [13,14]. However, several important questions remain unresolved. In particular, the structural evolution of glissile loop configurations during relaxation and its influence on loop mobility; the physical origin of the pronounced size dependence of loop diffusivity despite an apparently size-independent migration barrier [13]; and the atomic-scale mechanisms responsible for loop translation are not fully understood. In addition, establishing how the high intrinsic loop mobilities predicted by atomistic simulations relate to experimentally observed motion remains challenging, in part because quantitative measurements of individual loop motion at sufficiently high temporal resolution (e.g., on ms timescales) are scarce.

A small number of *in situ* transmission electron microscopy (TEM) studies have provided quantitative measurements of dislocation-loop motion. Shi *et al*. [15] directly investigated one-dimensional loop glide during ion irradiation of pure Ni at 773 K using a capture rate of 15 frames/s and measured glide velocities of 15–75 nm/s for the more mobile loops, with an average of 42 nm/s, while other loops exhibited substantially lower velocities of 3–4.8 nm/s. More recently, Hirst *et al*. [16] quantified dislocation-loop glide during *in situ* TEM annealing of neutron-irradiated Ti using images acquired at 1 frame/s and reported velocities of approximately 1.5–6.6 nm/s. Neither study compared the experimentally measured mobility with the intrinsic loop mobility predicted by atomistic simulations. Moreover, the temporal resolution of these measurements – at 67 ms and 1 s, respectively – limits their ability to resolve rapid fluctuations and short-lived transitions between mobile and pinned states. The resulting velocities are therefore time-averaged lower bounds. Thus, the simultaneous characterization of loop motion at substantially shorter timescales and comparison with atomistically predicted intrinsic mobilities remains an important gap.

In this work, we combine molecular dynamics simulations and high-speed (>1000 frames/s) *in situ* TEM observations to investigate the structure, energetics, and migration dynamics of self-interstitial clusters in FCC Ni. We first examine the formation and stability of disordered clusters, Frank loops, and perfect loops, and compare the energetics of Frank and perfect loop configurations over a broad range of cluster sizes. We then investigate the migration behavior of glissile perfect loops, including the effects of structural ordering, cluster size, and migration mechanism on loop mobility. Finally, we compare the intrinsic mobilities obtained from atomistic simulations with experimental loop motion observed at high speed and discuss the role of microstructural heterogeneities on loop transport in irradiated materials.

## 2. Methods

### *2.1. Simulation methodology*

We employed molecular dynamics (MD) simulations to study the behavior of self-interstitial atom (SIA) clusters in nickel. The simulations were performed using the open-source code LAMMPS [17]. Interatomic interactions in nickel were described using the embedded-atom method (EAM) potential developed by Bonny *et al.* [18] for Fe-Ni-Cr model alloys. This potential was chosen because it has been widely used for irradiation-defect simulations in Ni or Ni-containing FCC alloys and provides a suitable description of point defects and defect-cluster behavior in Ni [6,7]. Perfect FCC Ni simulation cells were first constructed with either 10,976 atoms in a $14a_0 \times 14a_0 \times 14a_0$ supercell or 87,808 atoms in a $28a_0 \times 28a_0 \times 28a_0$ supercell, where $a_0$ is the lattice parameter of FCC Ni. The cells were then equilibrated at the desired temperature using the NPT ensemble. SIA clusters were subsequently generated by introducing a prescribed number, $N$, of additional Ni atoms into the FCC simulation cell. The smaller cells were used primarily for small interstitial clusters (e.g., $N \leq 64$ for perfect loops simulations), while the larger cells were used for larger dislocation loops to reduce elastic interactions with periodic images. Periodic boundary conditions were applied in all three directions, and a time step of 1 fs was used.

In the initial cluster-generation simulations, interstitial atoms were randomly placed into the equilibrated simulation cell, followed by conjugate-gradient energy minimization and subsequent MD simulations in the NVT ensemble. After the finite-temperature MD runs, all saved configurations were again minimized using the conjugate-gradient algorithm before any further analysis. This minimize-anneal-minimize procedure was applied consistently to all simulations reported in this work. After relaxation, isolated SIAs formed dumbbell configurations with neighboring lattice atoms, with the preferred dumbbell orientation in Ni being along ⟨100⟩ [19]. Defect atoms were identified using Wigner-Seitz analysis implemented in OVITO [20], with a perfect FCC configuration at the same equilibrated density used as the reference lattice. Fig. 1(a) shows a representative simulation cell containing interstitial defects, where interstitial defect atoms, identified as atoms occupying Wigner-Seitz cells containing more than one atom, are shown in red. Because of their low concentration, the defect atoms are only faintly visible at the full-cell scale. Hence, defect configurations throughout this work are presented in the form shown in Fig. 1(b), where only the defect atoms identified by Wigner-Seitz analysis are shown and the remaining atoms are omitted for clarity. As the number of SIAs increased, elastic interactions drove the aggregation of individual dumbbells into larger clusters. Where appropriate, atomic structures were further characterized using the dislocation extraction algorithm (DXA) [21] developed by Stukowski *et al.* and implemented in OVITO.

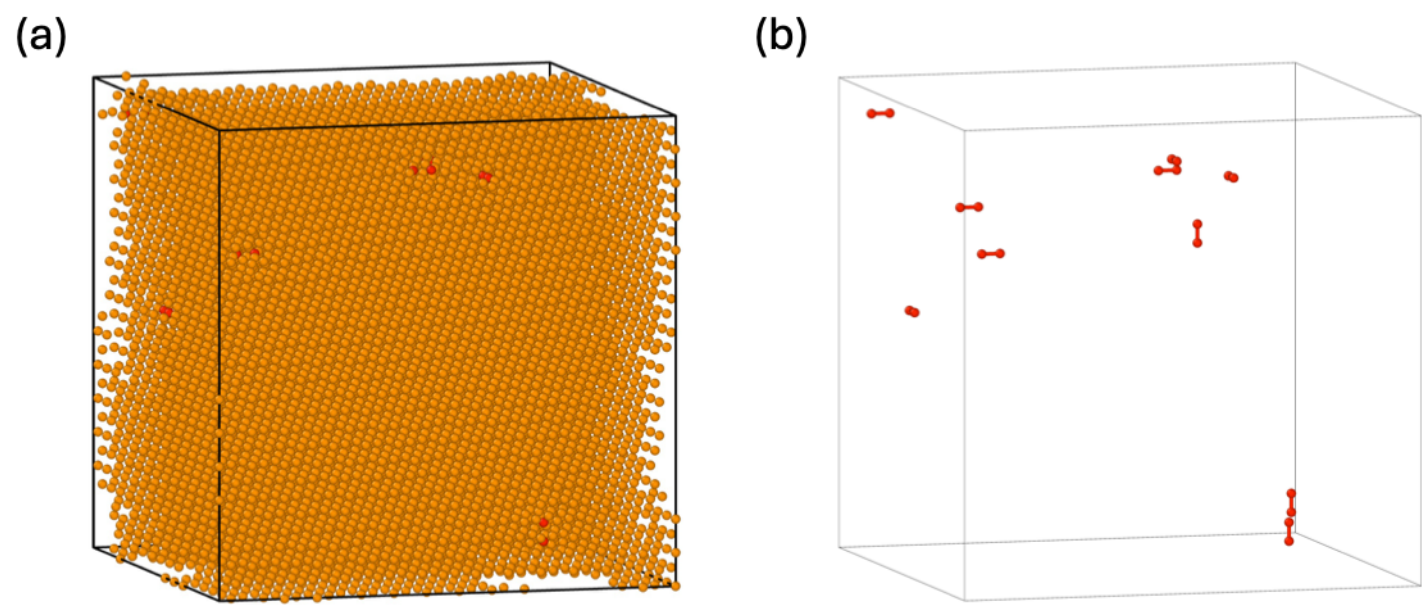

**Fig. 1.** Visualization of defect identification and representation in MD simulations. (a) Representative simulation cell containing randomly inserted SIAs after relaxation, with defect atoms identified by Wigner-Seitz analysis shown in red. Owing to the low defect concentration, the defect structure is difficult to discern at the full-cell scale. (b) Same configuration showing only the Wigner-Seitz defect atoms. The simulation box is retained here to illustrate the spatial scale and visualization convention. In subsequent figures, the view is cropped around the defect cluster, and the simulation box is omitted for clarity.

The interstitial clusters observed from the MD simulations were classified into three main categories: disordered clusters with mixed dumbbell orientations, sessile faulted Frank loops, and glissile ordered perfect (prismatic) loops. Frank loops were identified as faulted loops bounded by Frank partial dislocations and containing a {111} stacking fault. Perfect loops were identified as clusters with a dislocation loop enclosing the interstitials, predominantly aligned dumbbells, and rapid one-dimensional motion along the associated close-packed direction; these structures typically became more compact during MD annealing.

For larger $N$, spontaneous formation of a single well-defined Frank or perfect loop during MD became increasingly difficult because of kinetic limitations. For example, interstitial atoms could aggregate into multiple independent perfect loops that migrated in parallel rather than forming a single loop. Therefore, the recurring structural motifs identified from the MD simulations, namely the {111} stacking fault for Frank loops and compact aligned-dumbbell arrays for perfect loops, were used to construct both loop types directly in otherwise perfect FCC Ni simulation cells. For relatively small cluster sizes, the formation energies of these generated structures were compared with those of MD-relaxed loops to validate the construction method. After validation, the same procedure was extended to larger $N$ to compare Frank and perfect-loop energetics and to evaluate the size-dependent mobility of perfect loops.

The formation energy of a cluster containing $N$ self-interstitial atoms was calculated as

$$E_f(N) = E_{\text{def}} - (N_0 + N) \times \mu_{\text{Ni}} \tag{1}$$

where $E_{\text{def}}$ is the total energy of the defective simulation cell, $N_0$ is the number of atoms in the corresponding perfect FCC lattice, and $\mu_{\text{Ni}}$ is the cohesive energy per atom of FCC Ni calculated using the same interatomic potential. Because each defect cluster migrates as a single entity, its mobility was quantified by tracking the time-dependent displacement of the cluster center of mass (COM). The diffusion coefficient ($D$) was obtained from COM mean-squared displacement (MSD):

$$D = \frac{1}{2dt}\langle |r(t) - r(0)|^2 \rangle \tag{2}$$

where $r(t)$ is the COM position of the defect cluster at time $t$, $d$ is the dimensionality of the diffusion process, and $\langle \cdot \rangle$ denotes an ensemble average. For perfect loops undergoing one-dimensional migration along a $\langle 110 \rangle$ direction, $d$ = 1. The ensemble-averaged COM MSD, $\langle |r(t) - r(0)|^2 \rangle$, was calculated using a multiple-time-origin buffer-averaging procedure following Rapaport [22] and averaged over five independent simulations for each cluster size and temperature. The time interval $t$ used in Eq. (2) was chosen from the diffusive regime, where the log-log slope of the MSD versus time is approximately 1. Diffusion activation energy, $E_D$, was obtained by fitting the temperature-dependent $D$ to the Arrhenius form $D = D_0 \exp(-E_D/(k_B T))$, where $D_0$ is the pre-exponential factor, $k_B$ is the Boltzmann constant, and $T$ is the temperature in Kelvin.

To determine the microscopic mechanism of loop migration, atomic configurations were analyzed at short time intervals during MD trajectories. The displacement of individual dumbbells and the cluster's center of mass were monitored to distinguish rigid-body translation from localized rearrangement mechanisms. This analysis showed that perfect (prismatic) loops migrate through a relay mechanism, in which only a portion of the cluster advances during each activation step. The number of atoms participating in each activation event was determined by counting those undergoing coordinated displacement over a 1 ps interval, thereby allowing the size dependence of the migration mechanism to be correlated with the observed variation in the diffusion pre-exponential factor.

*2.2. Experimental details*

High-speed *in situ* TEM was used to monitor the thermally activated dynamics of nanoscale black-dot contrast features in Ni. The specimen was prepared from 4N (99.99%) pure Ni. An electron-transparent TEM lamella was prepared by focused ion beam (FIB) lift-out and thinning using a Thermo Scientific Helios 5 DualBeam FIB-SEM. The nanoscale tracked features in this study were introduced during $Ga^+$ FIB preparation of the lamella. Previous studies have shown that FIB-prepared metallic TEM specimens can contain black-spot damage associated with vacancy/interstitial clusters [23], and that FIB-cut Ni can contain Ga-rich regions, FIB-induced dislocation loops, and small black-dot features associated with FIB-generated point-defect clusters [24,25]. We therefore refer to the tracked features as FIB-induced black-dot defect clusters. Because the observations were performed under a single bright-field TEM (BF-TEM) two-beam condition, the Burgers vector, habit plane, and vacancy/interstitial character of individual defects could not be unambiguously determined.

The lamella was mounted on a custom Gatan 652 double-tilt heating holder and examined in an FEI Titan 80–200 aberration-corrected (S)TEM. The specimen was tilted from a [001] zone axis to a two-beam diffraction condition with g = (200), heated from room temperature to 500 °C at a rate of 9 °C/min, and held until thermal drift stabilized. After stabilization, a series of BF-TEM videos was recorded at 1100 frames/s using a Direct Electron Celeritas XS camera. Individual frames were extracted and smoothed using a Gaussian convolution with a standard deviation of 3 pixels, and the spatial trajectories of the FIB-induced black-dot defect clusters were manually tracked using ImageJ.

## 3. Results

### *3.1. Structural motifs formed by SIA clustering in nickel*

The self-interstitial atom (SIA) clusters formed during the MD simulations were found to evolve into three main classes of defect structures: disordered interstitial clusters, sessile Frank loops, and glissile perfect (prismatic) loops. Representative examples of these structures are shown in Fig. 2, and their characteristic structural and dynamical features are summarized in Table 1.

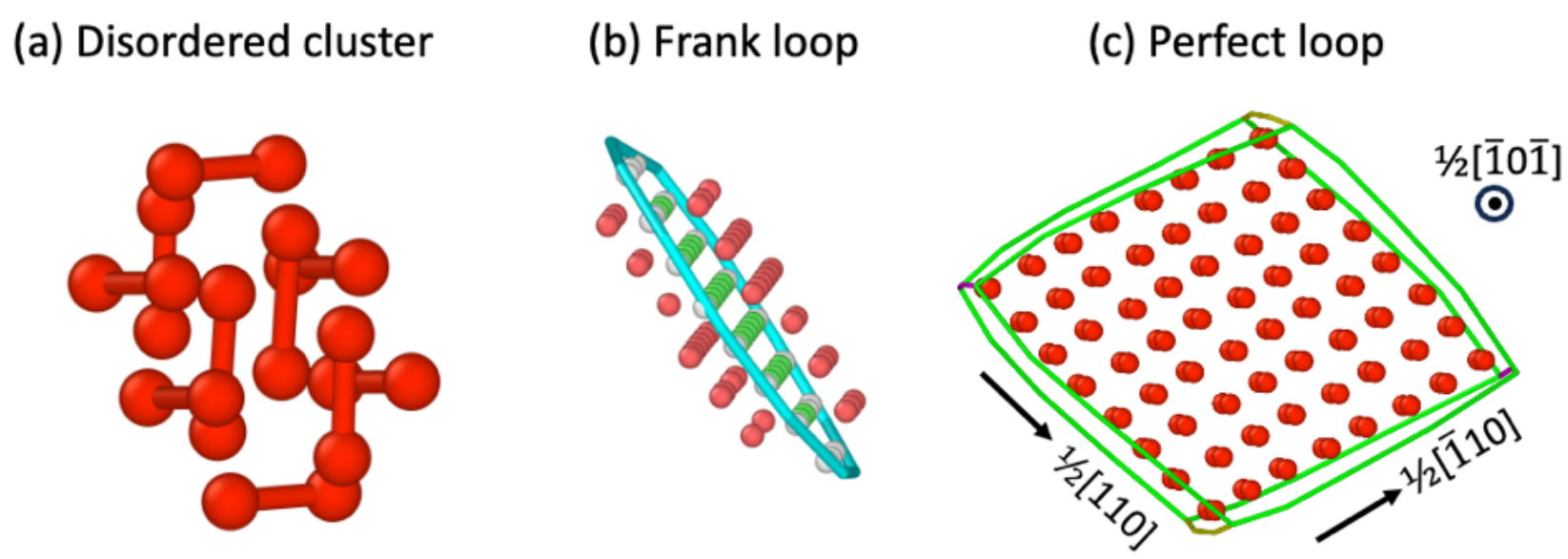

**Fig. 2.** Representative self-interstitial cluster structures observed in MD simulations of FCC Ni. (a) Disordered interstitial cluster composed of dumbbells with mixed orientations. (b) Sessile Frank loop containing a faulted {111} stacking fault bounded by a 1/3⟨111⟩ Frank partial dislocation (cyan line). Atoms are colored according to their local crystal structure: HCP (light red), FCC (green), and Other (white). (c) Glissile perfect (prismatic) loop consisting of an ordered arrangement of aligned dumbbells. The loop is bounded by a network of 1/6⟨112⟩ Shockley (green), 1/3⟨100⟩ Hirth (yellow), and 1/6⟨110⟩ stair-rod (purple) dislocations. The orientations of neighboring dumbbells within the loop are indicated. Only defect atoms identified by Wigner–Seitz analysis are shown.

For small cluster sizes, the interstitials commonly form disordered clusters composed of dumbbells with mixed orientations, although transient alignment can occasionally occur. For the limiting case of a single self-interstitial $N = 1$, the stable defect configuration is a ⟨100⟩ dumbbell, whose migration occurs through a sequence of translation and reorientation events [26,27]. A detailed analysis of the isolated dumbbell migration pathways and associated energy barriers is provided in Section S1 of the Supporting Information. The disordered clusters do not exhibit a well-defined loop character, and their mobility is intermediate compared with the loop structures discussed below. As the cluster size increases, two distinct loop morphologies are observed [13]. Frank loops are faulted dislocation loops bounded by Frank partial dislocations with Burgers vector $\boldsymbol{b} = 1/3\langle 111\rangle$ and containing a {111} stacking fault [28]. These loops are sessile on the MD timescales studied. In contrast, perfect loops consist of dumbbells aligned predominantly along a common close-packed direction and exhibit rapid one-dimensional migration along this direction [13]. Importantly, perfect loops were identified by the presence of a dislocation loop enclosing the interstitial cluster, as shown in Fig. 2(c), together with the common alignment of the dumbbells and the associated ⟨110⟩ migration direction; the dumbbell structures do not need to be perfectly compact immediately after formation.

**Table 1.** Classification of interstitial-cluster structures observed in MD simulations of FCC Ni. Clusters are classified as disordered clusters, sessile Frank loops, or glissile perfect (prismatic) loops according to their dumbbell orientations, dislocation character, stacking-fault structure, and mobility. Representative examples are shown in Fig. 2.

| Type | Dumbbell orientation | Observed size range | Mobility | Ground-state structure |
|---|---|---|---|---|
| Disordered cluster | Mixed ⟨1 0 0⟩ | $N \leq 10$ | Intermediate | $N < 9$ |
| Frank loop | Mixed ⟨1 0 0⟩ | $N \geq 14$ | Very low | Never |
| Perfect loop | Single ⟨1 1 0⟩ | $N \geq 6$ | Very high | $N \geq 9$ |

*3.2. Evolution of perfect loops toward compact ordered configurations*

Although perfect loops can form from aligned dumbbell arrangements, when formed from randomly placed evolving interstitials, their initial structures are often neither compact nor energetically optimized. During MD annealing, these loops undergo substantial structural reorganization while maintaining their overall perfect-loop character. Fig. 3 shows an example of an $N = 20$ SIA cluster, where an initially noncompact ordered cluster evolves toward a more compact arrangement during annealing. This evolution is accompanied by a decrease in formation energy and, as shown later, an increase in cluster mobility.

The formation of compact ordered configurations can nevertheless be kinetically limited. During the final stages of compact-cluster formation during MD, we observed cases in which only one dumbbell was shifted from its ideal position by a nearest-neighbor separation, and the barrier for this final rearrangement step was approximately 0.6 eV. This value is comparable to the barrier associated with isolated dumbbell reorientation/translation pathways shown in Fig. S1, indicating that local dumbbell rearrangements can control the rate at which compact structures form. For larger clusters, evolution toward the final compact shape may therefore require multiple such local rearrangements, making the ordering process progressively more difficult on MD timescales. The compact configurations therefore represent lower-energy states within the category of perfect loops, but their formation can require overcoming a sequence of kinetic barriers. For cluster sizes corresponding to perfect-square number of SIAs, such as $N =$ 25 and 49, the lowest-energy configurations from MD are nearly square arrays of aligned dumbbells. Once formed, these compact ordered structures remain stable over hundreds of nanoseconds of simulations and over the range of temperatures examined.

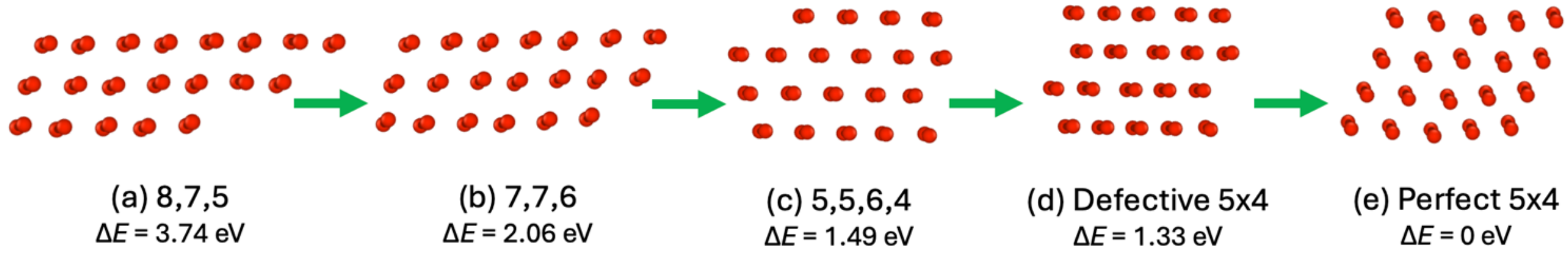

**Fig. 3.** Structural relaxation of an $N = 20$ perfect (prismatic) SIA loop during MD annealing. Initially noncompact perfect-loop configurations progressively reorganize into lower-energy structures through local dumbbell rearrangements while retaining their overall perfect-loop character. The labels indicate the number of dumbbells in successive rows, and $\Delta E$ denotes the formation energy relative to the lowest-energy configuration. Structural relaxation ultimately produces a compact ordered $5 \times 4$ perfect loop, which serves as the reference state for subsequent energetic and mobility analyses.

*3.3. Energetic comparison of Frank and perfect loops for varying N*

For large $N$, spontaneous formation of a single well-defined Frank or perfect loop becomes increasingly difficult in direct MD simulations because of kinetic limitations. Therefore, the recurring structural motifs identified from MD were used to construct both loop types directly over a wider range of cluster sizes to compare their energetics. Frank loops were generated by inserting an additional finite {111} atomic plane into an otherwise perfect FCC Ni crystal, with the inserted atoms selected from a compact region of a central {111} plane and shifted along the local [111] plane-normal direction before relaxation. Perfect loops were generated by inserting aligned dumbbells in the compact arrangement observed in the naturally formed MD clusters, using the dumbbell orientation and in-plane directions shown schematically in Fig. 2(c), before relaxation. This direct construction procedure was first validated by comparing the formation energies of generated loops with those of loop structures obtained directly from MD simulations at relatively small $N$. For the perfect loops, the compact aligned-dumbbell construction yielded highly reproducible relaxed structures, and the resulting formation energies exhibited negligible variation between independent realizations at a given cluster size. In contrast, the construction of Frank loops is not unique because different compact regions of the inserted {111} plane can be selected for a given $N$. To account for this variability, multiple candidate Frank-loop configurations were generated for each cluster size using different compact insertion geometries (including circular and various ellipsoidal regions). Following relaxation, a 10 ps annealing treatment, and a final relaxation, only configurations identified as Frank loops by DXA were retained. The lowest formation energy among the independent realizations was then taken as the representative Frank-loop energy for that cluster size.

As shown in Figs. 4(a) and (b), the generated Frank and perfect loop structures fall on the same energetic branches as the corresponding MD-relaxed loops. This agreement validates the synthetic construction procedure and enables the energetic comparison to be extended to larger cluster sizes. Fig. 4(c) compares the formation energies of generated Frank and perfect loops over the extended size range. Although the two branches remain close in total formation energy, the perfect loop is consistently lower in energy for all stable Frank-loop sizes examined. This trend is highlighted in Fig. 4(d), where the energy difference, $\Delta E = E_{\mathrm{Frank}} - E_{\mathrm{perfect}}$ remains positive over the entire range of $N$. Moreover, the energetic advantage of the perfect loop increases systematically with cluster size, consistent with the growing energetic penalty associated with the stacking fault contained within the Frank loop.

To assess the generality of these observations, analogous calculations were performed for Cu and Al (see Section S2 of the Supporting Information). We found that perfect loops in Cu are energetically favored over Frank loops across the full range of sizes examined, similar to the behavior observed in Ni. In contrast, Frank loops are more stable than perfect loops in Al. These results demonstrate that the relative stability of Frank and perfect loops is material dependent and is not universal across FCC metals. The trends obtained for Cu and Al are also consistent with recent first-principles calculations [29], which identified prismatic loops as the lowest-energy two-dimensional interstitial-loop family in Cu and Frank loops as the preferred two-dimensional loop structure in Al. However, those same calculations identified Frank loops as the lowest-energy two-dimensional interstitial-loop family in Ni for cluster sizes up to approximately 20 SIAs, whereas the present calculations find perfect loops to be energetically favored beginning at $N = 14$. This discrepancy may reflect differences in the interatomic potential, loop construction procedure, or the specific defect geometries sampled.

The results in Fig. 4 indicate that, within the present Ni potential, the perfect-loop branch is thermodynamically favored over the Frank-loop branch. However, transformation from a Frank loop to a perfect loop requires overcoming substantial kinetic barriers, allowing Frank loops to persist as long-lived metastable defects. This is illustrated by the $N = 20$ simulations, where both Frank and perfect loops are formed in different MD runs even though $E_{\mathrm{Frank}} > E_{\mathrm{perfect}}$. The Frank loop did not transform during 10 ns simulations at temperatures up to 1000 K, whereas transformation was observed at 1200 K through a multi-step pathway involving barriers of approximately 0.7–2 eV. Although MD simulations cannot directly determine Frank-loop lifetimes on experimental timescales, barriers near the upper end of this range are sufficiently large that thermally activated transformation can extend to seconds or longer at substantially lower temperatures. The strong temperature dependence is qualitatively consistent with the experiments of Chen *et al.* [10], who studied Frank-loop unfaulting in ion-irradiated 316L stainless steel and found that Frank-loop unfaulting was barely observed during irradiation at 350 °C but became pronounced at 400–450 °C, with no well-defined critical loop size. Taken together, these results suggest that the persistence of Frank loops in irradiated Ni is governed primarily by kinetic constraints rather than thermodynamic stability. If this is correct, the experimentally observed coexistence of Frank and perfect loops would most naturally be explained by kinetic trapping and defect-formation pathways rather than by the Frank–perfect energy difference alone.

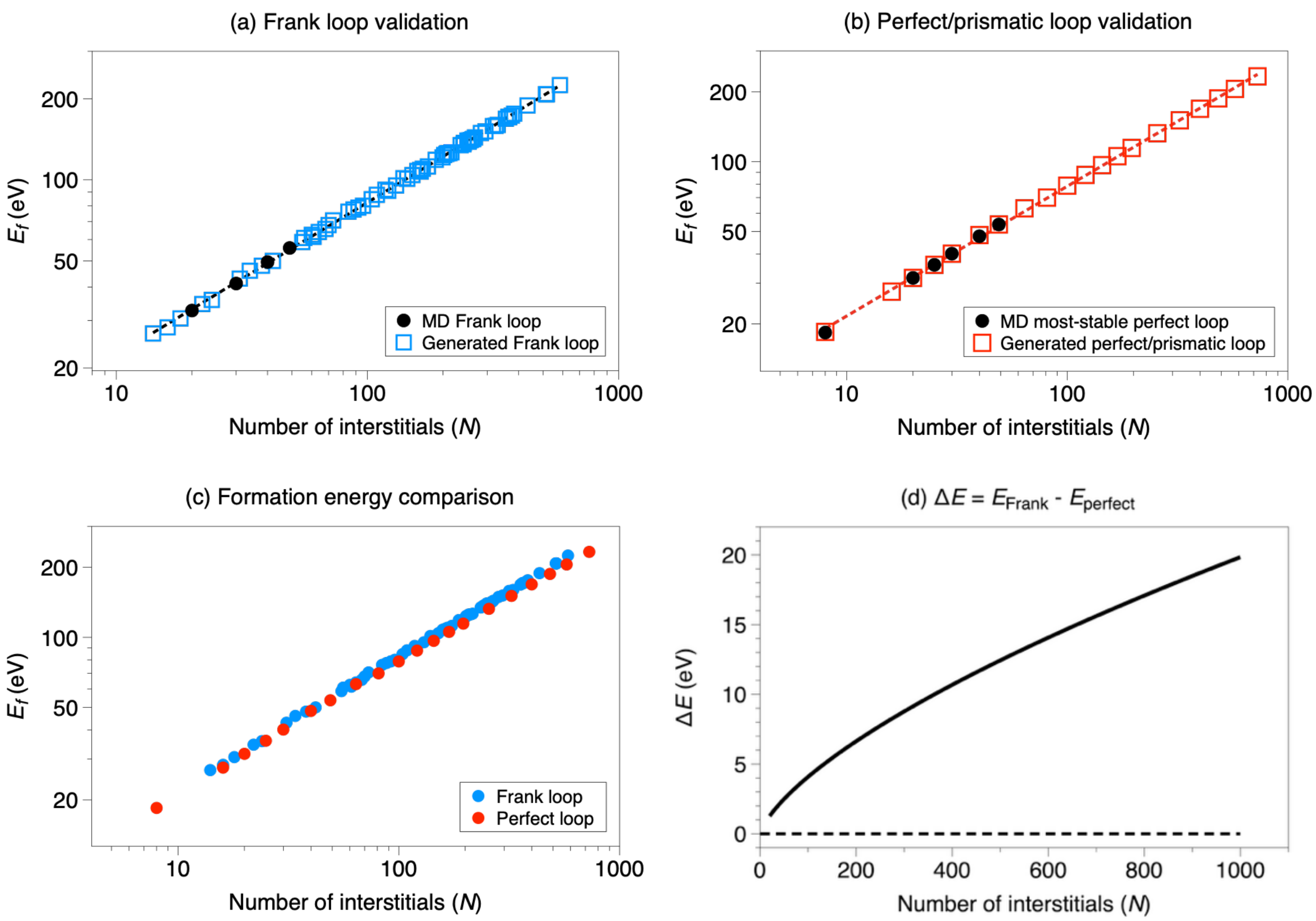

**Fig. 4.** Validation and energetic comparison of Frank and perfect (prismatic) SIA loops in Ni. (a,b) Formation energies of synthetically generated Frank and perfect loops compared with MD-relaxed structures at small cluster sizes, demonstrating the validity of the loop-construction procedure. (c) Formation energies of generated Frank and perfect loops over an extended range of cluster sizes. (d) Energy difference, $\Delta E = E_{\mathrm{Frank}} - E_{\mathrm{perfect}}$, demonstrating that perfect loops are thermodynamically favored over Frank loops throughout the range of stable Frank-loop sizes examined.

### *3.4. Diffusion of small disordered interstitial clusters*

Before considering the mobility of well-defined perfect loops, we first examined the diffusion of small disordered interstitial clusters. The migration of an isolated SIA provides a useful reference state, and its diffusion behavior and activation energy are consistent with previous atomistic studies of Ni. Details of the isolated SIA migration pathway and associated NEB analysis are provided in Section S1 of the Supporting Information. For small clusters with $N < 6$, the defects remain predominantly disordered, similar to the structure shown in Fig. 2(a), and their diffusion can be analyzed without persistent loop formation. These clusters migrate through a combination of dumbbell translation and reorientation events, resulting in more complex motion than the one-dimensional glide observed for perfect loops. At $N = 6$, the simulations begin to show intermittent formation of perfect-loop-like configurations, which introduces rapid one-dimensional migration. This intermittent structural transformation makes the diffusion of larger "disordered" clusters difficult to define unambiguously; therefore, the diffusion analysis of disordered clusters was restricted to $N \leq 5$. Fig. S3 in the Supporting Information shows the time evolution of the cluster MSD for $N = 1$–5 at 900 K, where a clear diffusive regime is observed, with log-log slope of MSD versus time close to unity. The corresponding diffusion coefficients are shown in Fig. 5 and indicate that, over this size range and at this temperature, the mobility of disordered interstitial clusters generally decreases with increasing cluster size.

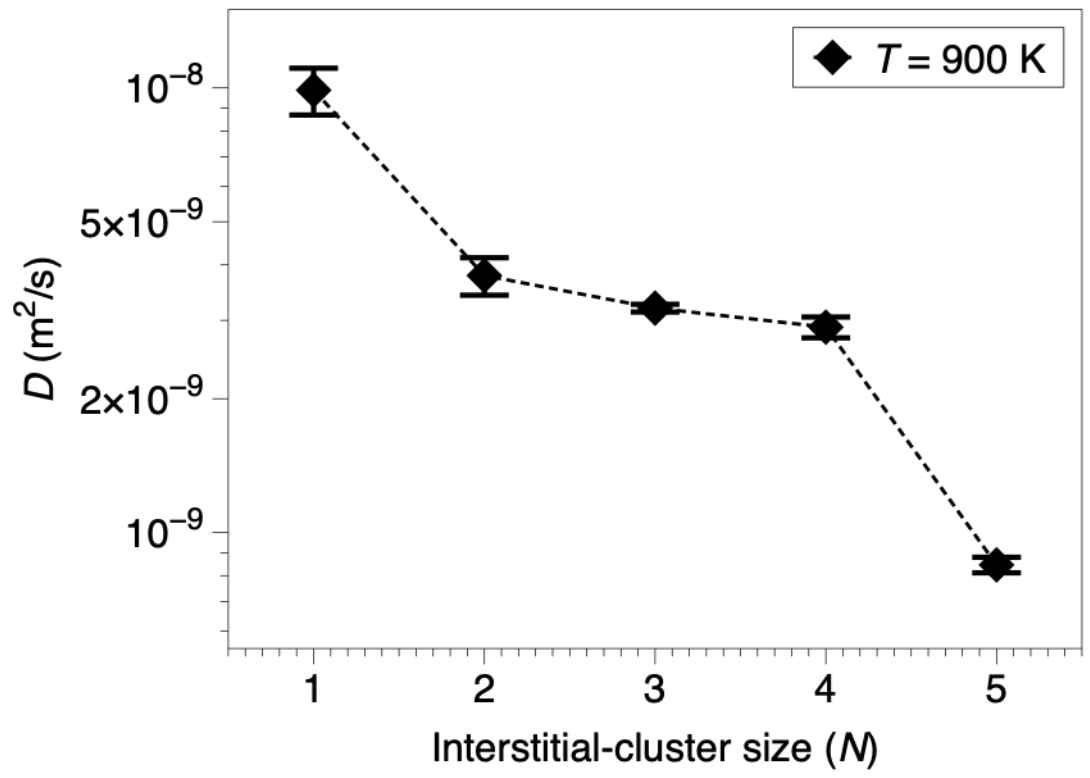

**Fig. 5.** Diffusion coefficients of small disordered interstitial clusters in Ni. For $N \leq 5$, the clusters remain predominantly disordered and migrate through a combination of dumbbell translation and reorientation events. The diffusion coefficient shows an overall decrease with increasing cluster size within this regime. Error bars represent the standard error of the mean obtained from five independent MD simulations.

### *3.5. Ordering and compactness enhance perfect-loop mobility*

The mobility of perfect loops depends strongly on their degree of structural ordering and compactness. During annealing, initially noncompact ordered clusters evolve toward more compact configurations. To quantify this relationship, we analyzed five $N = 20$ perfect-loop configurations with different degrees of compactness, shown in Fig. 3, using their formation energy relative to the most compact $5 \times 4$ ordered cluster as a measure of structural stability. As shown in Fig. 6, lower-energy configurations exhibit higher diffusivity at 400 K and smaller diffusion activation energies. The diffusivity is reported at 400 K because the least stable configurations readily transform into more compact structures at higher temperatures, making it difficult to measure their mobility without simultaneous structural evolution. The activation energies were therefore estimated from diffusion coefficients obtained over just 100–200 K where each configuration remained identifiable. These results show that structural relaxation within the perfect-loop branch is accompanied by enhanced one-dimensional mobility: noncompact ordered loops occupy higher-energy, less-mobile states, whereas compact ordered loops are both energetically favored and more diffusive. This trend suggests a hierarchy within the perfect-loop branch, in which increasing compactness simultaneously lowers the formation energy and enhances one-dimensional diffusion.

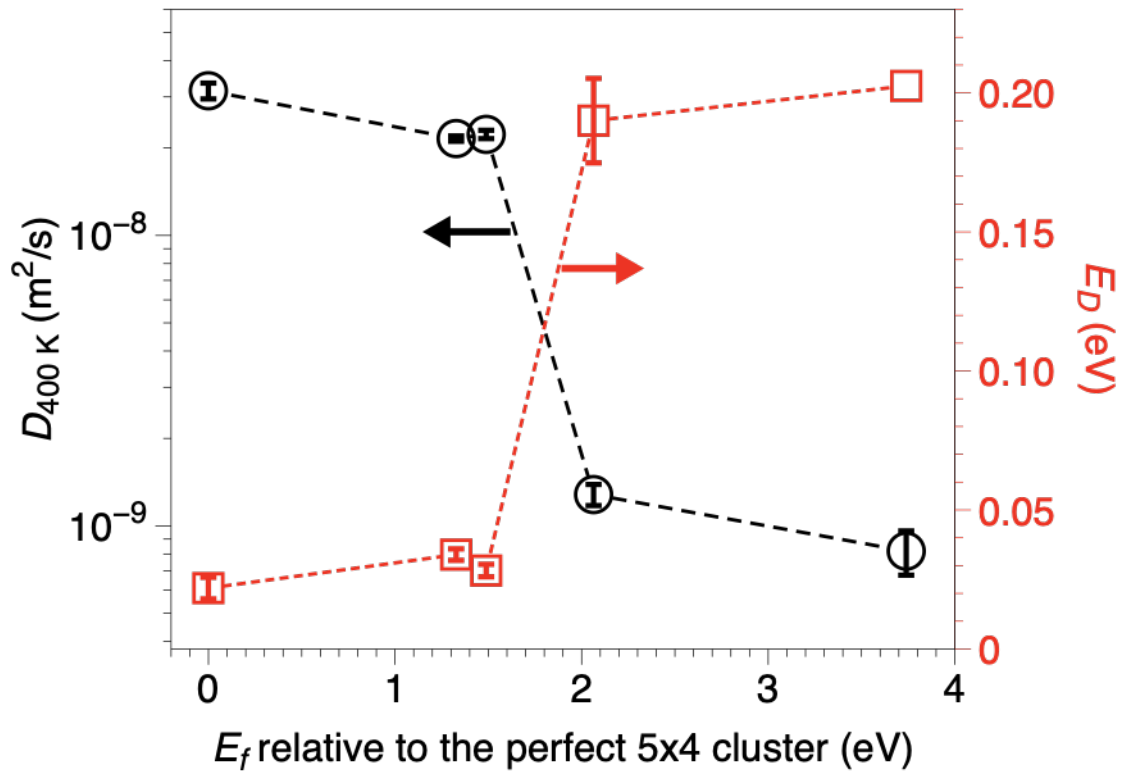

**Fig. 6.** Effect of structural ordering and compactness on perfect-loop mobility. Five $N = 20$ perfect-loop configurations with different degrees of compactness were characterized by their formation energies relative to the most compact (5 × 4) ordered configuration. More compact, lower-energy configurations exhibit higher diffusivities at 400 K and lower diffusion activation energies, demonstrating that structural relaxation within the perfect-loop family enhances one-dimensional mobility. Error bars in the diffusion coefficients represent the standard error of the mean from five independent MD simulations, while error bars in $E_D$ represent the standard error of the activation energy obtained from the Arrhenius fit.

### *3.6. Size-dependent mobility and relay-wise migration of perfect loops*

The diffusion coefficient of compact, energetically stable perfect loops was evaluated for cluster sizes ranging from $N$ = 16 to 400. Across this range, the loops exhibit rapid one-dimensional migration along the common dumbbell orientation. Fig. S4 shows the time evolution of the COM MSD at different temperatures for the two limiting sizes, $N$ =16 and 400, illustrating that both clusters reach a diffusive regime. In addition, Fig. S5 shows that the diffusion coefficient extracted from the MSD remains reasonably stable over a broad range of time intervals, supporting the use of the selected time interval for evaluating $D$. The MSDs were calculated using the maximum possible time origins within each trajectory and averaged over five independent simulations at each temperature. The diffusion coefficients for different $N$ and temperatures are shown in Fig. 7(a). At a given temperature, the diffusion coefficient decreases with increasing cluster size. Arrhenius fits to the diffusion coefficients, shown as dotted lines in Fig. 7(a), indicate that the diffusion activation energy ($E_D$) is very small, approximately 0.02 eV, and nearly independent of cluster size. The size dependence of the diffusion coefficient is therefore controlled primarily by the pre-exponential factor, $D_0$, which decreases with increasing $N$ and follows an approximate power-law scaling of $D_0 \propto N^{-0.54}$, as shown in Fig. 7(b).

This weak size dependence of the activation energy is consistent with earlier atomistic studies of one-dimensional SIA-cluster migration. Osetsky *et al.* [30] investigated cluster diffusion in FCC Cu and BCC Fe and reported nearly size-independent activation energies of approximately 0.02–0.03 eV for the jump frequency of ⟨111⟩ crowdion clusters in Fe up to $N \leq 91$, and for ⟨110⟩ crowdion-cluster motion in Cu over the more limited range $4 \leq N \leq 25$. Since the MSD did not exhibit a sufficiently clean linear-time dependence in those simulations, they focused primarily on jump-frequency analysis rather than on diffusion coefficients [30]. In a later study, Osetsky *et al.* [6] estimated diffusion coefficients for one-dimensionally migrating SIA clusters using trajectory-time decomposition and trajectory-jump decomposition, in which MD trajectories were divided into fixed-time or fixed-jump segments. Their trajectory-time decomposition is conceptually equivalent to the multiple-time-origin MSD averaging used here, apart from possible differences in whether overlapping or non-overlapping time intervals are used. Their diffusion coefficients and activation barrier for $N$ = 9 are in excellent agreement with the expected results from Fig. 7(a). However, they reported larger barriers for $N$ = 19 and 36, approximately 0.078 eV and 0.166 eV, respectively. One possible reason for this difference is that those clusters may continue to undergo structural evolution during the simulations; as discussed above, larger clusters may require longer times to reach their most stable configurations. In that case, the extracted diffusion coefficients would reflect both cluster migration and ongoing configurational relaxation, rather than the mobility of a fixed, compact loop structure alone.

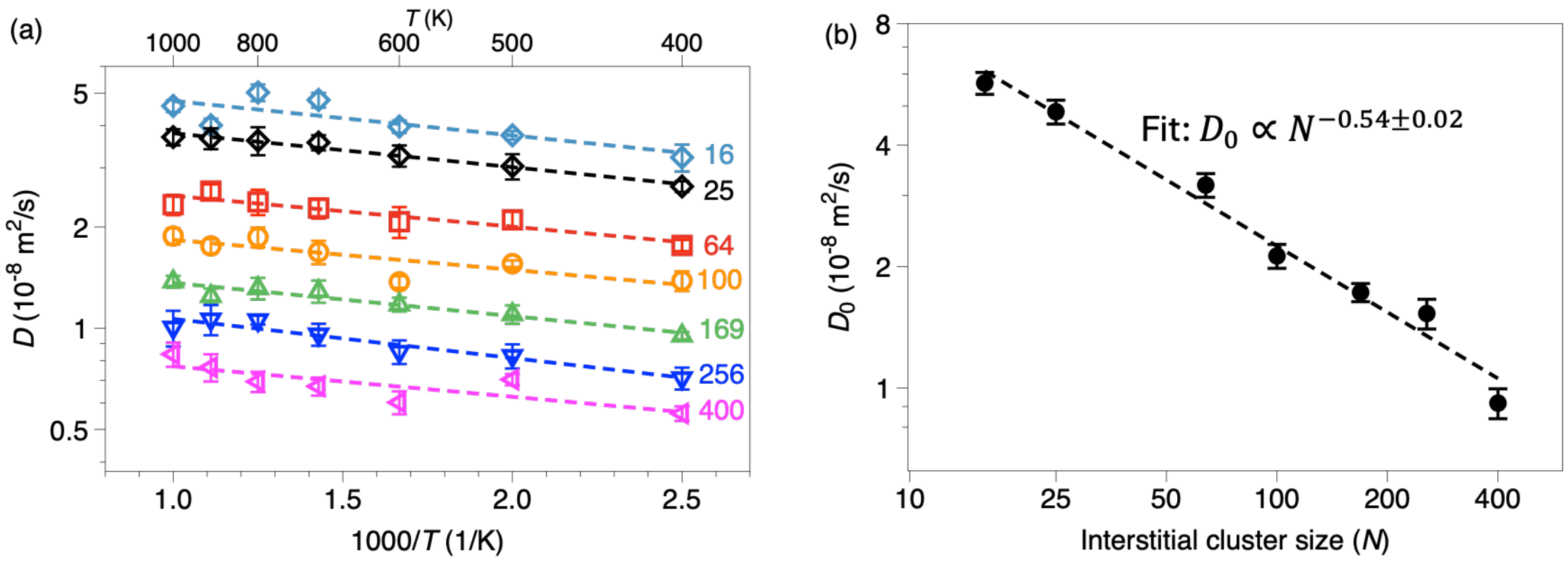

**Fig. 7.** Size-dependent diffusion of compact perfect (prismatic) loops. (a) Arrhenius plot of diffusion coefficients for compact perfect loops with cluster sizes ranging from $N = 16$ to 400 (denoted in text at the termination of each trendline). Dotted lines indicate Arrhenius fits. The diffusion activation energy is small (~0.02 eV) and nearly independent of cluster size. Error bars represent the standard error of the mean obtained from five independent MD simulations. (b) Diffusion prefactor, $D_0$, as a function of cluster size, showing an approximate power-law decrease with increasing $N$. Standard errors on the $D_0$'s and the power-law exponent are included.

We next examine the origin of the cluster-size dependence of $D_0$. In Ref. [30], the jump-frequency pre-factor for one-dimensionally migrating SIA clusters was reported to scale with cluster size as $N^{-s}$, with $s \approx 0.6$ in both Fe and Cu. The authors suggested that this trend reflected a reduced probability of correlated crowdion motion in larger clusters, although this interpretation was not supported by a quantitative analysis of the atoms participating in each migration event. To understand the scaling observed in the present work, we consider the transition-state-theory expression

$$D_0 = a^2 g f x_D \nu^*,$$

where $a$ is the hop distance, $g$ and $f$ are geometric and correlation factors, respectively, $x_D$ is the concentration of the diffusion-mediating defect, and $\nu^*$ is the attempt frequency for cluster migration [31]. If migration occurs through collective translation of the entire loop, with all SIAs moving cooperatively in a single step, the dominant size dependence of $D_0$ is expected to arise from $\nu^*$. Harmonic transition-state theory predicts that the attempt frequency scales inversely with the square root of the effective mass of the translating loop. Since the effective mass of the translating loop scales linearly with the number of SIAs, this gives $\nu^* \propto N^{-0.5}$, and consequently $D_0 \propto N^{-0.5}$, close to the scaling observed here. To test this prediction, we performed additional MD simulations of an $N = 100$ cluster in which the Ni atomic mass was varied from one-quarter to four times its nominal value while keeping the interatomic potential unchanged. The resulting diffusion coefficient (Fig. S6) followed $D \propto m^{-0.52}$, in excellent agreement with the predicted $D \propto m^{-1/2}$ dependence. This agreement supports the interpretation that the observed cluster-size dependence of $D_0$ originates primarily from the inverse-square-root mass dependence of the attempt frequency.

However, a direct analysis of the MD trajectories shows that the loops do not undergo rigid-body translation. Instead, migration occurs predominantly through a row-wise relay mechanism, in which only a subset of the aligned dumbbell rows advances during each activation step. In this picture, the relevant moving mass is characterized by the average number of participating atoms per migration event, $\langle N_{move} \rangle$, suggesting that the size dependence of $D_0$ should be governed by $\langle N_{move} \rangle$ rather than the total loop size, $N$.

To quantify $\langle N_{\mathrm{move}} \rangle$ for different cluster sizes, we analyzed derived dumbbell trajectories, in which each trajectory point represents the mean position of the two atoms sharing a Wigner-Seitz cell. Such trajectories can be constructed unambiguously here because the dumbbells migrate along well-defined one-dimensional paths without crossing one another. The dumbbell-center positions were recorded every 0.2 ps at 400 K and smoothed over five frames (encompassing 1 ps interval) to suppress rapid forward-backward fluctuations. Displacements were then evaluated over successive 1 ps intervals and projected onto the glide direction. A dumbbell was counted as participating in an activated event when its projected displacement exceeded 2.2 Å, approximately the nearest-neighbor spacing. The number of participating dumbbells was averaged over all nonzero migration events to obtain $\langle N_{\mathrm{move}} \rangle$, and this is shown in Fig. 8. This analysis shows that $\langle N_{\mathrm{move}} \rangle$ increases systematically with cluster size and follows an approximate power-law scaling, $\langle N_{\mathrm{move}} \rangle \propto N^{0.82}$. The scaling remains slightly sublinear across the range of displacement thresholds and time intervals examined, with fitted exponents ranging from approximately 0.80 to 0.89 (Fig. S7).

If the effective moving mass scales with $\langle N_{\mathrm{move}} \rangle$, the harmonic argument predicts $D_0 \propto N^{-0.41}$. Thus, the growth of the participating segment with loop size can account for most of the observed $D_0 \propto N^{-0.54}$ dependence. The remaining size dependence may reflect deviations from the harmonic approximation, as well as correlation effects not captured by the simple mass-scaling picture, including forward-backward correlations between successive relay events. These results indicate that the decrease in loop mobility with increasing size originates primarily from the increasing number of atoms involved in each migration event rather than from changes in the migration barrier itself.

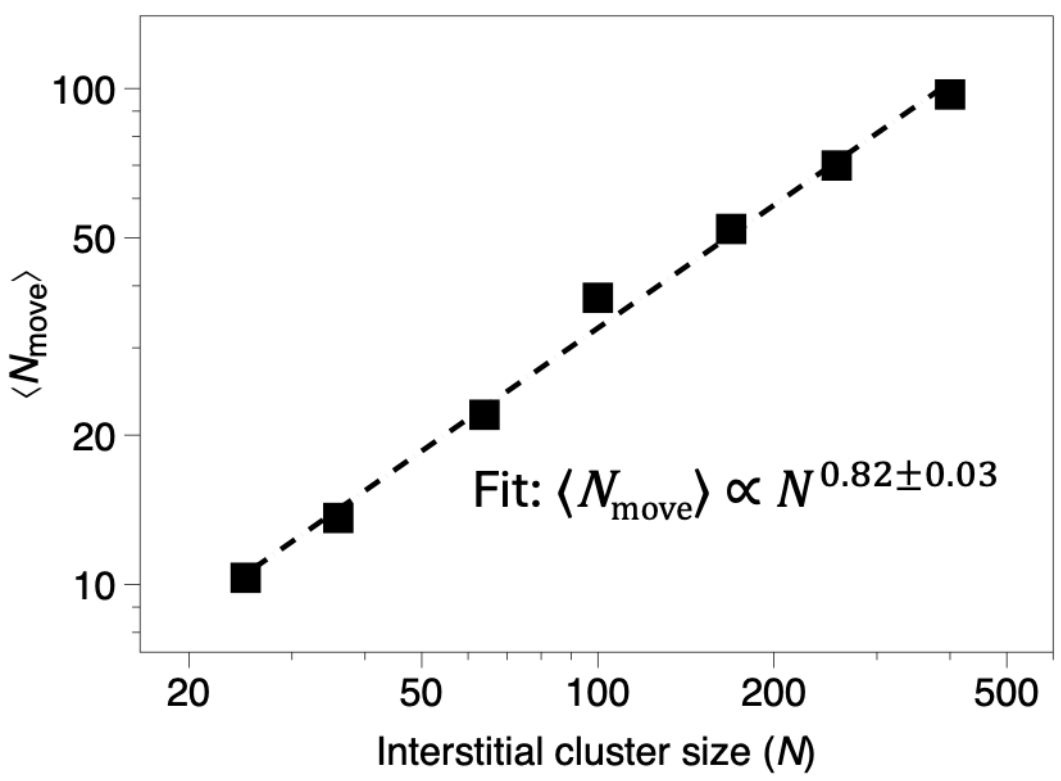

**Fig. 8.** Size dependence of the number of dumbbells participating in perfect-loop migration events. The average number of participating dumbbells, $\langle N_{\mathrm{move}} \rangle$, was obtained from derived dumbbell-center trajectories by counting dumbbells whose projected displacement along the glide direction exceeded one nearest-neighbor spacing over a 1 ps interval. $\langle N_{\mathrm{move}} \rangle$ increases systematically with cluster size and follows an approximate power-law scaling, indicating that larger loops migrate through increasingly collective row-wise relay events. The corresponding increase in effective moving mass accounts for a substantial fraction of the observed decrease in the diffusion prefactor with increasing $N$.

*3.7. Experimental observation of 1D dislocation loop motion*

Finally, the intrinsic mobility obtained from MD simulations is compared with sub-millisecond TEM observations of defect-loop motion in FIB-irradiated Ni. One-dimensional motion of interstitial-defect clusters formed during irradiation has previously been observed using *in situ* TEM in both BCC and FCC materials [8,32]. Here, in irradiated Ni, the loops were identified to predominantly migrate along a $\langle 110 \rangle$ direction, as illustrated by the TEM images in Fig. 9 and the representative trajectory shown in Fig. 10(a) for a defect loop of diameter ~20 nm. The observed motion is highly intermittent rather than continuous. The projected displacement along the migration direction, shown in Fig. 10(b), consists of periods of nearly constant displacement, referred to here as *pinning segments*, separated by *mobile segments* during which the loop advances approximately linearly with time. The perpendicular motion is nearly random, providing an estimate of the noise in the defect position assessment. For the trajectory shown in Fig. 10(a), three distinct pinning segments were identified. During the mobile segments, the measured loop velocity ranged from roughly 270 to 410 nm/s. The maximum image-to-image velocity, determined from displacement between successive frames that are separated by 0.9 ms, is ~$10^4$ nm/s. The presence of both pinned and mobile segments suggests that migrating loops undergo repeated trapping and release events due to interactions with defects and heterogeneities in the irradiated microstructure, consistent with the observations of Ono *et al.* [14] in FCC Cu. Similar behavior was observed in two additional trajectories of the same defect analyzed in this work (not shown). These trajectories span less than 0.1 s and contain pinning intervals of approximately 2–15 ms. During the mobile segments, the measured velocities are approximately 3000–4000 nm/s, substantially higher than those observed for the ~20 nm loop in Fig. 9. The defect in these additional trajectories is a ~6 nm black-spot cluster, whereas the defect shown in Fig. 9 is a small loop rather than a black-spot cluster. For comparison, previous in situ TEM observations in Ni [15] reported loop velocities of only ~3–4 nm/s for a defect of size comparable to that shown in Fig. 9, substantially lower than the velocities measured here.

Direct comparison between experimental observations and the intrinsic loop mobility obtained from MD simulations is not straightforward because the two do not exhibit the same types of motion. To facilitate this comparison, we consider simple models that relate the observed loop trajectories to an effective diffusivity, while recognizing the assumptions inherent in each approach. First, we approximate the mobile segments as one-dimensional random-walk diffusion. Although the mobile segments are approximately linear and therefore do not display the characteristic $\sqrt{t}$ scaling of unbiased diffusion, their displacement scale can be mapped onto an equivalent one-dimensional random walk to define an apparent coarse-grained diffusivity. Under the one-dimensional random-walk assumption, the displacement distribution is Gaussian with variance $2D_{\mathrm{RW}}t$, where $D_{\mathrm{RW}}$ is the random-walk diffusion coefficient and $t$ is the observation time. The corresponding mean absolute displacement is $\langle|x|\rangle = \sqrt{4D_{\mathrm{RW}}t/\pi}$, giving an apparent velocity $\langle v\rangle = \sqrt{4D_{\mathrm{RW}}/\pi t}$, or

$$D_{\mathrm{RW}} = {}^{\pi t}\!/_{4}\,\langle v\rangle^2. \tag{3}$$

Substituting the average mobile-segment velocity, $\langle v\rangle \approx 300$ nm/s and average mobile-segment duration, $t \approx 0.1$ s, yields an effective diffusion coefficient of $7 \times 10^{-15}$ m$^2$/s. This value is approximately five orders of magnitude smaller than the intrinsic diffusivity predicted by the MD simulations for a 20 nm perfect loop at 500 °C. Thus, if the experimentally observed motion is interpreted as a one-dimensional random-walk process, the effective loop mobility is dramatically reduced relative to that of a free loop in an ideal crystal. This large discrepancy suggests that additional physical processes, such as trapping at microstructural defects, impurity drag, or other local heterogeneities, strongly suppress loop transport in the TEM specimen.

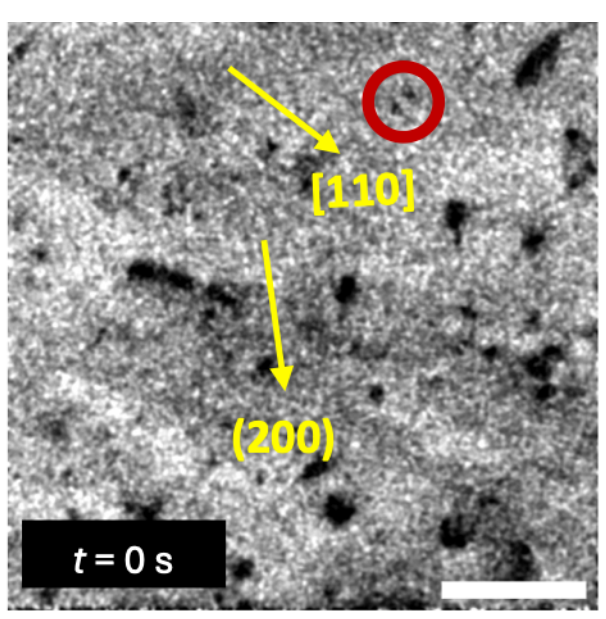

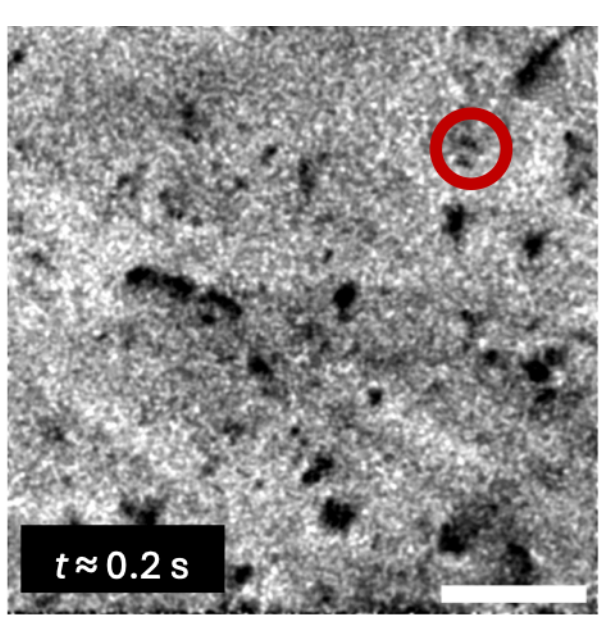

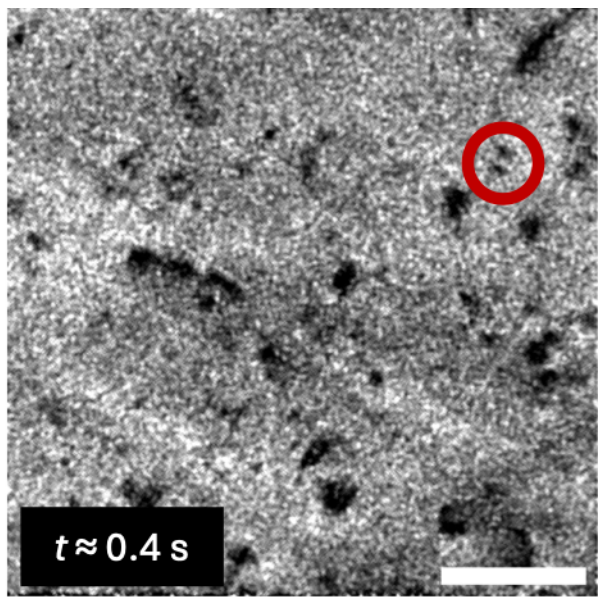

**Fig. 9.** Experimental observation of one-dimensional defect-loop motion in irradiated Ni. Sequential TEM images showing the motion of a nanoscale defect loop along a [110] direction at $t = 0$, ≈0.2, ≈0.4 s. The red circles identify the tracked defect loop, and the yellow lines indicate the [110] migration direction and the (200) direction. Scale bars, 80 nm.

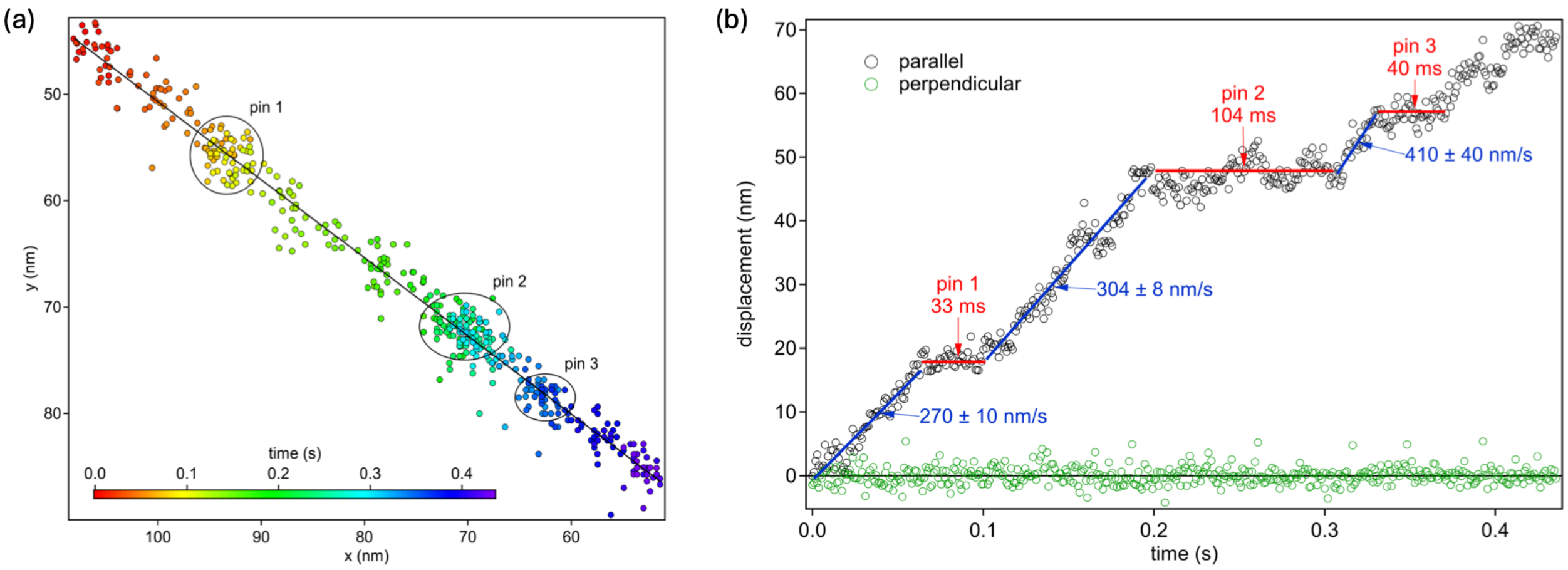

**Fig. 10.** Trajectory and displacement analysis of the one-dimensional loop motion. (a) Representative loop trajectory obtained from manually extracted loop positions. Colors indicate time progression, and the circled regions identify the three pinning sites. (b) Displacement parallel and perpendicular to the migration direction as a function of time, illustrating repeated trapping and release events.

As a second model, we treat the linear displacement observed during the mobile segments in Fig. 10(b) as drift-like motion rather than diffusive transport. Under this assumption, the observed drift velocity is used to infer an equivalent diffusivity through the Einstein relation. In the presence of a local driving force, the loop velocity is given by $v = MF_{local}$, where $M$ is the effective loop mobility and $F_{local}$ is the local driving force. Using the Einstein relation, $M = D/k_BT$, the diffusion coefficient corresponding to the observed drift velocity is

$$D_{\mathrm{drift}} = {}^{vk_BT}/_{F_{local}}. \tag{4}$$

Because the local driving force is not independently known, the trajectory in Fig. 10 alone cannot uniquely determine $D_{\mathrm{drift}}$; however, an approximate estimate can be obtained by adopting a representative value of the local driving force, $F_{\mathrm{local}}$. The observation of TEM defect-loop trajectories exhibiting different migration directions, together with partially reversible motion, indicates that the driving forces acting on the loops are local rather than uniform, consistent with migration through a heterogeneous energy landscape rather than under a single macroscopic applied stress. Assuming a representative local stress, $\sigma$, of approximately 100 MPa (~0.05% of the bulk modulus of Ni), which is reasonable for a FIB-prepared TEM lamella because of residual stresses, implanted Ga gradients, surface damage, and local foil curvature, the corresponding $F_{local}$ can be estimated. For a dislocation loop subjected to a local resolved stress $\sigma$, the Peach–Koehler force per unit length is $f_{PK} \sim \sigma b$, where $b$ is the Burgers vector. The total driving force can therefore be estimated as $F_{local} \sim \sigma b L_{eff}$, where $L_{eff}$ is the effective loop length participating in motion. Taking $b$ = 0.25 nm for FCC Ni, and approximating $L_{eff}$ by the loop circumference, $L_{eff} {\sim} \pi d$, with $d \sim$ 20 nm, and a representative local stress $\sigma$ of ~100 MPa gives $F_{local} \sim 1.6 \times 10^{-9}$ N. Using this value in Eq. (4) gives $D_{\mathrm{drift}} \sim 2 \times 10^{-18}$ m$^2$/s. Again, this estimated diffusivity is several orders of magnitude lower than the intrinsic diffusivity predicted by the molecular dynamics. Because $D_{\mathrm{drift}}$ is inversely proportional to $F_{\mathrm{local}}$, lower assumed values of the local stress $\sigma$ or of the effective participating loop length $L_{eff}$ would correspondingly increase the inferred diffusion coefficient.

Finally, as a third model to estimate the loop diffusivity, we focus on the fluctuations during the pinning segments. The trajectories in Fig. 10(b) indicate that, during the three pinning intervals, the defect loop remains confined within a finite region while continuing to undergo substantial fluctuations. We therefore model the pinned loop as a one-dimensional Brownian particle confined within a reflecting interval of characteristic width $W$. Analysis of the three pinning intervals gives a mean absolute displacement between consecutive TEM frames of $\langle|\Delta x|\rangle \approx 1.44$ nm, while the spatial extent of the fluctuations gives a characteristic confinement width of $W \approx 5$ nm (more details in the Supporting Information, Section S7). For a Brownian particle confined within a reflecting interval, the normalized mean absolute displacement is governed by the dimensionless parameter $\alpha = D\Delta t/W^2$, such that

$$\frac{\langle|\Delta x|\rangle}{W} = f\left(\frac{D\Delta t}{W^2}\right)$$

where $\Delta t = 0.9$ ms is the time between the observations and $f(\alpha)$ is determined analytically from the reflecting-boundary propagator (see Section S7). The normalized mean absolute displacement, $\langle|\Delta x|\rangle/W$, increases monotonically from 0, as $D \rightarrow 0$, to 1/3 in the limit of sufficiently rapid diffusion, where consecutive positions become statistically independent within the confinement region. Therefore, for specified $W$ and $\Delta t$, a measured value between these limits corresponds to a unique effective confined diffusivity $D_{\text{confined}}$ within the reflecting-well model. For $W \approx 5$ nm, this decorrelated limit corresponds to $\langle|\Delta x|\rangle = W/3 \approx 1.67$ nm. The experimentally measured value of 1.44 nm is close to this limit, indicating substantial intrawell mobility while retaining finite positional correlation between successive frames. For the experimentally measured ratio $\langle|\Delta x|\rangle/W = 1.44/5 = 0.288$, the analytical confined-diffusion relation gives $\alpha \approx 0.152$, as illustrated in Fig. S8. Brownian-dynamics simulations independently reproduce this result (Section S7). Using $W = 5$ nm and $\Delta t = 0.9$ ms in $\alpha = D\Delta t/W^2$ then yields an effective confined diffusivity of $D_{\text{confined}} \approx 4.2 \times 10^{-15}$ m$^2$ s$^{-1}$. Within this model, the apparently stationary portions of the trajectory do not represent an immobile defect loop. Rather, the loop remains highly mobile within a nanoscale pinning region while its long-range translation is temporarily suppressed by the local energy landscape. The full derivation of the analytical relation between $\langle|\Delta x|\rangle/W$ and $D\Delta t/W^2$, together with its independent validation using Brownian dynamics, is provided in Section S7.

Table 2 lists the diffusivities from the MD simulations and the TEM analyses. The TEM-based estimates are substantially lower than the intrinsic mobility predicted by MD for an isolated loop in a perfect crystal. The random-walk and confined diffusion models give diffusivities of $7 \times 10^{-15}$ and $4.2 \times 10^{-15}$ m$^2$/s, respectively. Considering that these values are obtained from different portions of the trajectory and using different physical models, their agreement within a factor of two is notable. Both are approximately five orders of magnitude below the MD prediction of $\sim 10^{-9}$ m$^2$/s. This large reduction is consistent with the intermittent nature of the experimental trajectories, which exhibit repeated pinning, changes in migration direction, and partially reversible motion, and suggests that loop transport in the irradiated TEM specimen is strongly influenced by the heterogeneous local environment rather than being governed solely by the intrinsic migration kinetics captured by the ideal-crystal MD simulations.

The large discrepancy between the $D$ values extracted from TEM and MD strongly suggests that the physics governing defect-loop motion in the experiments differs substantially from that in the ideal single-crystal MD simulations. While this is expected due to loop pinning, we find that the discrepancy persists even when we focus on portions of the motion from which the effects of the directly observed pinning have been removed, as in the three models discussed above. There are at least two natural sources of this discrepancy. The first is that additional pinning may occur on timescales shorter than the 0.9 ms temporal resolution of the experiment. If the defect pins and unpins many times during a single imaging interval, what we interpret as free motion would in fact represent an average over unresolved pinning and migration events. Another closely related idea is impurity drag associated with a Cottrell atmosphere [33,34]. In this mechanism, impurities (in this case, implanted Ga) are attracted to the loop and move along with it, effectively reducing its mobility because impurity diffusion is expected to be much slower than the intrinsic loop motion. Determining the origin of the remaining discrepancy between the diffusivities extracted from the TEM experiments and MD simulations is an important direction for future work and will be necessary to fully understand and model defect-loop motion.

**Table 2.** Summary of diffusion coefficients estimated from MD simulations and from TEM observations using the random-walk, driving-force, and pinning analyses. The key assumptions and input parameters used in each estimate are also listed.

| Estimation method | Key inputs/assumptions | Estimated diffusivity, $D$ |
|---|---|---|
| MD | Perfect loop, $d$ = 20 nm; $T$ = 500 °C | $D_{MD} \sim 10^{-9}$ m$^2$/s |
| TEM (mobile regime, random-walk model) | $v \approx 300$ nm/s; mobile segment duration $t \approx 0.1$ s; random-walk assumption | $D_{RW} \approx 7 \times 10^{-15}$ m$^2$/s |
| TEM (mobile regime, driving-force model) | $v \approx 300$ nm/s; $\sigma \approx 100$ MPa; $b$ = 0.25 nm; $d \approx 20$ nm; $L_{eff} \approx \pi d$ | $D_{drift} \approx 2 \times 10^{-18}$ m$^2$/s |
| TEM (pinning regime, confined-diffusion model) | $W \approx 5$ nm; $\langle|\Delta x|\rangle \approx 1.44$ nm; $\Delta t \approx 0.9$ ms | $D_{confined} \approx 4.2 \times 10^{-15}$ m$^2$/s |

## 4. Conclusions

Self-interstitial atom (SIA) clustering is a key early process in radiation damage accumulation in FCC metals because interstitial clusters evolve into dislocation loops that govern defect transport, recombination, hardening, and the long-term microstructural evolution of irradiated materials. Although both sessile Frank loops $\boldsymbol{b} = 1/3\langle 111\rangle$ and glissile perfect (prismatic) loops $\boldsymbol{b} = 1/2\langle 110\rangle$ are observed in irradiated FCC Ni, the structural evolution of perfect loops, the origins of their size-dependent mobility, and their energetic relationship to Frank loops have remained incompletely understood. By combining molecular dynamics (MD) simulations with high-speed *in situ* TEM observations, we investigated the structure, energetics, and migration dynamics of SIA clusters in FCC Ni. The main conclusions are:

1. Perfect loops undergo substantial structural relaxation after formation. Initially noncompact configurations progressively reorganize into compact ordered dumbbell arrays, accompanied by a significant increase in mobility. For cluster sizes corresponding to perfect square number of SIAs (e.g., $N$ = 25 and 49), the lowest-energy structures from MD are nearly perfect square arrays of dumbbells that remain stable over hundreds of nanoseconds across the temperatures examined. Recognition of this recurring structural motif enabled the direct construction of stable perfect loops over a much broader range than is accessible through spontaneous MD nucleation.

2. Direct construction and relaxation of Frank and perfect loops over a wide range of cluster sizes show that perfect loops are energetically favored for all examined sizes above $N$ = 14, with the energetic advantage increasing systematically with cluster size. This trend is consistent with the growing stacking-fault energy penalty associated with the Frank-loop configuration. Despite this thermodynamic preference, substantial kinetic barriers hinder Frank-loop unfaulting, allowing Frank loops to persist as long-lived metastable defects. The experimentally observed coexistence of Frank and perfect loops [10] is therefore attributed to kinetic trapping rather than comparable thermodynamic stability. An important open question is what microscopic mechanisms during defect-cluster nucleation and early evolution determine whether a cluster initially forms as a Frank or a perfect loop.

3. The diffusion coefficient of perfect loops evaluated for $N$ = 16–400 exhibits a migration barrier that is both small (~0.02 eV) and essentially independent of cluster size. The strong size dependence of loop mobility arises primarily from the diffusion prefactor, which decreases approximately as $D_0 \propto N^{-0.54}$.

4. Perfect-loop migration does not occur through rigid-body translation of the entire cluster. Instead, loops advance through a row-wise relay mechanism in which only a subset of dumbbells participates in each migration event. The number of participating atoms increases systematically with cluster size, scaling approximately as $N^{0.82}$, and accounts for much of the observed reduction in the diffusion prefactor through its influence on the effective attempt frequency.
5. High-speed *in situ* TEM observations reveal intermittent migration of nanoscale defect loops in irradiated Ni. The coexistence of mobile and pinned segments, changes in migration direction, and partially reversible trajectories indicate that loop migration is governed by local driving forces, trapping, and impurity interactions rather than exhibiting free one-dimensional diffusion. Multiple analyses of the mobile and pinned portions of the trajectories yield effective diffusivities in the range of $10^{-18}$ to $10^{-14}$ $m^2/s$, approximately five to nine orders of magnitude below the intrinsic MD prediction of $10^{-9}$ $m^2/s$. These large discrepancies are consistent with additional pinning and impurity effects still to be determined.

Together, these results establish the structural evolution, thermodynamic stability, and migration mechanisms of ideal interstitial loops in FCC Ni across a broad range of cluster sizes. The resulting migration barriers, mobility scaling laws, and loop-stability criteria can provide physically grounded inputs for mesoscale models of radiation-induced microstructural evolution. However, the large discrepancies between the MD and the experimental observations suggest that loop motion is massively reduced by pinning and perhaps other forms of drag. These processes must be understood and integrated into models before quantitative loop kinetics can be fully understood, modeled, and used to quantitatively interpret experiments or guide microstructural modeling.

## CRediT authorship contribution statement

AA and DM conceived the MD simulation work. AA performed the MD simulations and analyzed the results. IM and HL prepared the TEM samples, performed the TEM measurements, and analyzed the TEM trajectory data under the supervision of PMV and KGF. AA developed the models used to estimate defect-loop mobilities from the experimental observations, with input from DM. AA wrote the original draft of the manuscript. All authors contributed to reviewing and editing the manuscript. KGF, PMV, and DM supervised the work and secured funding.

## Data availability

The data supporting the findings of this study will be made openly available in the Figshare repository upon publication at [URL will be added].

## Declaration of Competing Interest

The authors declare no competing financial interests.

## Acknowledgements

This work was primarily supported by the U.S. Department of Energy, Office of Science, Office of Basic Energy Sciences under Award Number DE-SC0025213. The authors gratefully acknowledge Prof. Julie Tucker at Oregon State University for synthesizing and providing the 4N-pure Ni material used in this work. TEM imaging used facilities in the Wisconsin Centers for Nanoscale Technology, supported in part by the University of Wisconsin Materials Research Science and Engineering Center (DMR-2309000). This work used the TACC's Stampede3 at the University of Texas at Austin through allocation TG-MAT240071, from the Advanced Cyberinfrastructure Coordination Ecosystem: Services & Support (ACCESS) program, which is supported by the National Science Foundation (NSF) grants #2138259, #2138286, #2138307,

#2137603, and #2138296. The authors are also grateful to the Center for High Throughput Computing (CHTC) at UW–Madison for the computing resources.

## References

[1] A.F. Rowcliffe, L.K. Mansur, D.T. Hoelzer, R.K. Nanstad, Perspectives on radiation effects in nickel-base alloys for applications in advanced reactors, Journal of Nuclear Materials 392 (2009) 341–352. https://doi.org/10.1016/J.JNUCMAT.2009.03.023.

[2] M. Griffiths, Ni-Based Alloys for Reactor Internals and Steam Generator Applications, Structural Alloys for Nuclear Energy Applications (2019) 349–409. https://doi.org/10.1016/B978-0-12-397046-6.00009-5.

[3] P.J. Maziasz, Overview of microstructural evolution in neutron-irradiated austenitic stainless steels, Journal of Nuclear Materials 205 (1993) 118–145. https://doi.org/10.1016/0022-3115(93)90077-C.

[4] G.S. Was, Fundamentals of radiation materials science: Metals and alloys, Fundamentals of Radiation Materials Science: Metals and Alloys (2007) 1–827. https://doi.org/10.1007/978-3-540-49472-0/SAVE-RESEARCH.

[5] D.S. Aidhy, C. Lu, K. Jin, H. Bei, Y. Zhang, L. Wang, W.J. Weber, Point defect evolution in Ni, NiFe and NiCr alloys from atomistic simulations and irradiation experiments, Acta Mater. 99 (2015) 69–76. https://doi.org/10.1016/J.ACTAMAT.2015.08.007.

[6] Y.N. Osetsky, A. V. Barashev, Y. Zhang, On the mobility of defect clusters and their effect on microstructure evolution in fcc Ni under irradiation, Materialia (Oxf). 4 (2018) 139–146. https://doi.org/10.1016/J.MTLA.2018.09.028.

[7] C. Lu, T. Yang, L. Niu, Q. Peng, K. Jin, M.L. Crespillo, G. Velisa, H. Xue, F. Zhang, P. Xiu, Y. Zhang, F. Gao, H. Bei, W.J. Weber, L. Wang, Interstitial migration behavior and defect evolution in ion irradiated pure nickel and Ni-xFe binary alloys, Journal of Nuclear Materials 509 (2018) 237–244. https://doi.org/10.1016/J.JNUCMAT.2018.07.006.

[8] C. Lu, L. Niu, N. Chen, K. Jin, T. Yang, P. Xiu, Y. Zhang, F. Gao, H. Bei, S. Shi, M.R. He, I.M. Robertson, W.J. Weber, L. Wang, Enhancing radiation tolerance by controlling defect mobility and migration pathways in multicomponent single-phase alloys, Nature Communications 2016 7:1 7 (2016) 13564-. https://doi.org/10.1038/ncomms13564.

[9] C. Chen, Y. Wang, J. Hou, J. Song, Dislocation Loop Transformation in Metals: Computational Studies, Theoretical Prediction and Future Perspectives, Acc. Mater. Res. 6 (2025) 473–483. https://doi.org/10.1021/ACCOUNTSMR.4C00296/ASSET/IMAGES/LARGE/MR4C00296_0009.JPEG.

[10] D. Chen, K. Murakami, K. Dohi, K. Nishida, Z. Li, N. Sekimura, The effects of loop size on the unfaulting of Frank loops in heavy ion irradiation, Journal of Nuclear Materials 529 (2020) 151942. https://doi.org/10.1016/J.JNUCMAT.2019.151942.

[11] C. Dai, Q. Wang, P. Saidi, B. Langelier, C.D. Judge, M.R. Daymond, M.A. Mattucci, Atomistic structure and thermal stability of dislocation loops, stacking fault tetrahedra, and voids in face-centered cubic Fe, Journal of Nuclear Materials 563 (2022) 153636. https://doi.org/10.1016/J.JNUCMAT.2022.153636.

[12] P. Xiu, H. Bei, Y. Zhang, L. Wang, K.G. Field, STEM Characterization of Dislocation Loops in Irradiated FCC Alloys, Journal of Nuclear Materials 544 (2021) 152658. https://doi.org/10.1016/J.JNUCMAT.2020.152658.

[13] Y.N. Osetsky, D.J. Bacon, A. Serra, B.N. Singh, S.I. Golubov, Stability and mobility of defect clusters and dislocation loops in metals, Journal of Nuclear Materials 276 (2000) 65–77. https://doi.org/10.1016/S0022-3115(99)00170-1.

[14] K. Ono, M. Miyamoto, K. Yamahaku, In-situ observation of the dynamic behavior of cascade defect clusters formed by irradiation with high-energy self-ions at 50 K in Cu, Journal of Nuclear Materials 511 (2018) 122–127. https://doi.org/10.1016/J.JNUCMAT.2018.09.001.

[15] S. Shi, H. Bei, I.M. Robertson, Impact of alloy composition on one-dimensional glide of small dislocation loops in concentrated solid solution alloys, Materials Science and Engineering: A 700 (2017) 617–621. https://doi.org/10.1016/J.MSEA.2017.05.049.

[16] C.A. Hirst, B. Kombaiah, I.L. Steigerwald, K.G. Field, M.P. Short, In situ TEM annealing of neutron-irradiated Ti reveals a two-stage mechanism for elevated temperature radiation damage recovery, Scr. Mater. 271 (2026) 117001. https://doi.org/10.1016/J.SCRIPTAMAT.2025.117001.

[17] S. Plimpton, Fast Parallel Algorithms for Short-Range Molecular Dynamics, J. Comput. Phys. 117 (1995) 1–19. https://doi.org/https://doi.org/10.1006/jcph.1995.1039.

[18] G. Bonny, D. Terentyev, R.C. Pasianot, S. Poncé, A. Bakaev, Interatomic potential to study plasticity in stainless steels: the FeNiCr model alloy, Model. Simul. Mat. Sci. Eng. 19 (2011) 085008. https://doi.org/10.1088/0965-0393/19/8/085008.

[19] P. Zhao, Y. Shimomura, Molecular dynamics calculations of properties of the self-interstitials in copper and nickel, Comput. Mater. Sci. 14 (1999) 84–90. https://doi.org/10.1016/S0927-0256(98)00077-9.

[20] A. Stukowski, Visualization and analysis of atomistic simulation data with OVITO-the Open Visualization Tool, Model. Simul. Mat. Sci. Eng. 18 (2010) 015012. https://doi.org/10.1088/0965-0393/18/1/015012.

[21] A. Stukowski, V. V. Bulatov, A. Arsenlis, Automated identification and indexing of dislocations in crystal interfaces, Model. Simul. Mat. Sci. Eng. 20 (2012) 085007. https://doi.org/10.1088/0965-0393/20/8/085007.

[22] D.C. Rapaport, The Art of Molecular Dynamics Simulation, 2nd ed., Cambridge University Press, 2004. https://doi.org/10.1017/CBO9780511816581.

[23] D.J. Edwards, A. Schemer-Kohrn, M. Olszta, R. Prabhakaran, Y. Zhu, J. Wang, J. Haag, O. El Atwani, T.G. Lach, M. Toloczko, Understanding and removing FIB artifacts in metallic TEM samples using flash electropolishing, Journal of Nuclear Materials 606 (2025) 155618. https://doi.org/10.1016/J.JNUCMAT.2025.155618.

[24] V. Samaeeaghmiyoni, H. Idrissi, J. Groten, R. Schwaiger, D. Schryvers, Quantitative in-situ TEM nanotensile testing of single crystal Ni facilitated by a new sample preparation approach, Micron 94 (2017) 66–73. https://doi.org/10.1016/J.MICRON.2016.12.005.

[25] V. Samaee, R. Gatti, B. Devincre, T. Pardoen, D. Schryvers, H. Idrissi, Dislocation driven nanosample plasticity: new insights from quantitative in-situ TEM tensile testing, Scientific Reports 2018 8:1 8 (2018) 12012-. https://doi.org/10.1038/s41598-018-30639-8.

[26] P.W. Ma, S.L. Dudarev, Nonuniversal structure of point defects in face-centered cubic metals, Phys. Rev. Mater. 5 (2021) 013601. https://doi.org/10.1103/PHYSREVMATERIALS.5.013601/FIGURES/27/MEDIUM.

[27] S. Bukkuru, U. Bhardwaj, K.S. Rao, A.D.P. Rao, M. Warrier, M.C. Valsakumar, Kinetics of self-interstitial migration in bcc and fcc transition metals, Mater. Res. Express 5 (2018) 035513. https://doi.org/10.1088/2053-1591/AAB418.

[28] D. Hull, D.J. Bacon, Introduction to Dislocations, Fifth Edition, Introduction to Dislocations, Fifth Edition (2011) 1–257. https://doi.org/10.1016/C2009-0-64358-0.

[29] A.M. Goryaeva, C. Domain, A. Chartier, A. Dézaphie, T.D. Swinburne, K. Ma, M. Loyer-Prost, J. Creuze, M.C. Marinica, Compact A15 Frank-Kasper nano-phases at the origin of dislocation loops in face-centred cubic metals, Nature Communications 2023 14:1 14 (2023) 3003-. https://doi.org/10.1038/s41467-023-38729-6.

[30] Y.N. Osetsky, D.J. Bacon, A. Serra, B.N. Singh, S.I. Golubov, One-dimensional atomic transport by clusters of self-interstitial atoms in iron and copper, Philosophical Magazine 83 (2003) 61–91. https://doi.org/10.1080/0141861021000016793.

[31] A. Van der Ven, G. Ceder, M. Asta, P.D. Tepesch, First-principles theory of ionic diffusion with nondilute carriers, Phys. Rev. B Condens. Matter Mater. Phys. 64 (2001). https://doi.org/10.1103/PhysRevB.64.184307.

[32] K. Arakawa, K. Ono, M. Isshiki, K. Mimura, M. Uchikoshi, H. Mori, Observation of the one-dimensional diffusion of nanometer-sized dislocation loops, Science (1979). 318 (2007) 956–959. https://doi.org/10.1126/SCIENCE.1145386/SUPPL_FILE/ARAKAWA.SOM.PDF.

[33] C. Nowak, X.W. Zhou, R.B. Sills, Validating continuum theory for Cottrell atmosphere solute drag by molecular dynamics simulations, J. Mech. Phys. Solids 183 (2024) 105514. https://doi.org/10.1016/J.JMPS.2023.105514.

[34] Y. Mishin, J.W. Cahn, Thermodynamics of Cottrell atmospheres tested by atomistic simulations, Acta Mater. 117 (2016) 197–206. https://doi.org/10.1016/J.ACTAMAT.2016.07.013.

### *S1. Motion of a single self-interstitial in nickel*

As a baseline for the analyses presented in this work, we first examine the migration of a single self-interstitial atom (SIA) in nickel, whose behavior is well established. The stable configuration of a single SIA in FCC Ni is a ⟨100⟩ dumbbell [1]. To characterize the elementary migration mechanism of this defect, we calculated the energy barriers associated with migration between neighboring dumbbell positions using the nudged elastic band (NEB) method in LAMMPS. During the migration process, both translational motion and changes in dumbbell orientation were considered. Fig. S1(a) illustrates the migration pathways examined. Four possible combinations of translation and reorientation were evaluated, corresponding to different initial (shown in gray) and final dumbbell orientations. The solid and dotted lines connect the positions of the same atom in the initial and final configurations and are shown only to guide the eye. The resulting energy profiles are shown in Fig. S1(b). The migration barrier depends strongly on the orientation pathway, with the lowest-energy pathway corresponding to a coupled translation–reorientation process [1,2], having an activation barrier of ~0.33 eV. Alternative pathways exhibit substantially higher barriers, ranging from approximately 0.66–0.84 eV.

To determine which migration pathway controls long-range diffusion, molecular dynamics simulations of a single SIA in Ni were performed over a range of temperatures. The resulting diffusion coefficients are shown in Fig. S1(c). Arrhenius analysis yields a diffusion activation energy of 0.33 eV, in excellent agreement with the lowest barrier obtained from the NEB calculations and with the results of Osetsky *et al.* [3]. This agreement indicates that long-range diffusion of isolated SIAs proceeds predominantly through the lowest-energy translation–reorientation pathway [2]. The correspondence between the NEB barrier and the diffusion activation energy further suggests that dumbbell reorientation is an intrinsic component of SIA migration over the temperature range examined.

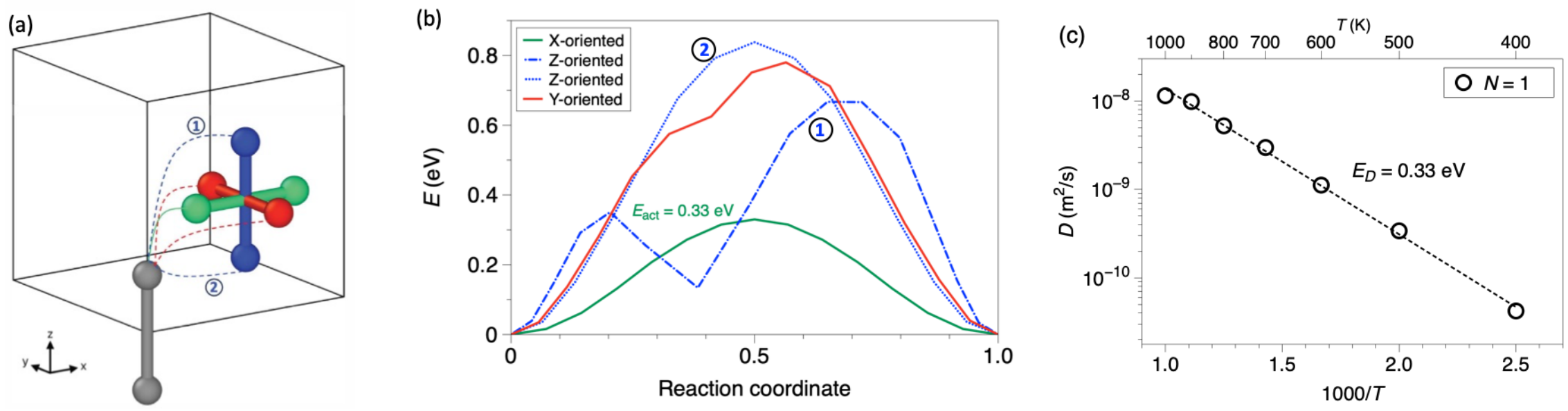

**Fig. S1.** Migration of a single self-interstitial atom in FCC Ni. (a) Schematic illustration of the migration pathways considered between neighboring ⟨100⟩ dumbbell configurations. Dashed lines indicate representative translation–reorientation pathways. (b) NEB energy profiles for the migration pathways shown in (a). The lowest-energy pathway exhibits an activation barrier of approximately 0.33 eV, whereas alternative pathways have significantly higher barriers. (c) Arrhenius plot of the diffusion coefficient for a single SIA obtained from MD simulations. The diffusion activation energy extracted from the Arrhenius fit, $E_D$ = 0.33 eV, agrees closely with the lowest NEB barrier, indicating that long-range diffusion is controlled by the coupled translation–reorientation pathway.

### *S2. Energetics of Frank and perfect loops in copper and aluminum*

To examine whether the energetic trends observed for Ni loops are general across FCC metals, the same synthetic loop-construction and relaxation procedure was applied to Cu and Al. Formation energies were calculated using the embedded-atom method (EAM) potential developed for both Cu and Al by Mendelev *et al.* [4], employing the same methodology used for Ni. Frank and perfect loops were generated over a range of cluster sizes, relaxed, and their formation energies compared.

Fig. S2 shows the resulting formation energies. Although the Cu and Al datasets exhibit somewhat greater variability than those obtained for Ni, reflecting the less extensive structural optimization performed for these comparative calculations, the qualitative energetic trends are unambiguous. In Cu, the perfect (prismatic) loop remains energetically favored over the Frank loop throughout the size range examined, consistent with the behavior observed in Ni. In contrast, Frank loops are consistently lower in energy than perfect loops in Al. These findings therefore demonstrate that the relative stability of Frank and perfect loops is strongly material dependent. They are also consistent with recent first-principles calculations [5], which identified prismatic loops as the preferred two-dimensional interstitial-loop family in Cu and Frank loops as the preferred two-dimensional loop family in Al.

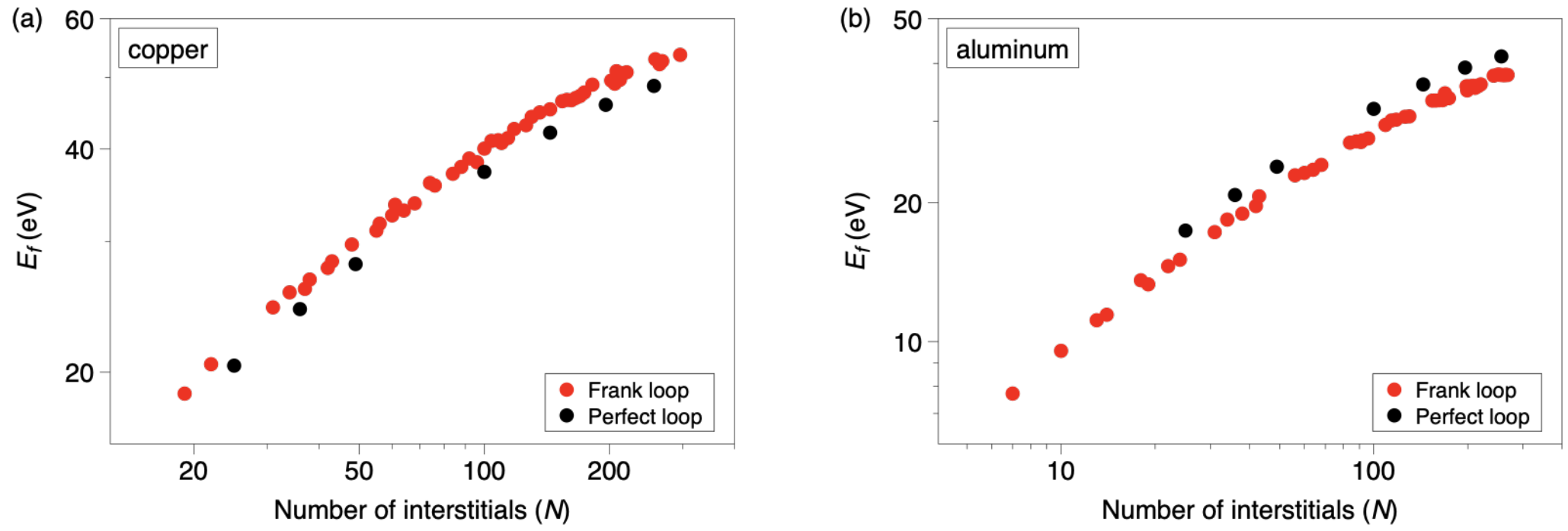

**Fig. S2.** Formation energies of generated Frank and perfect self-interstitial loops in (a) Cu and (b) Al as a function of cluster size $N$. Calculations were performed using the same synthetic loop-construction and relaxation procedure employed for Ni. Perfect loops are energetically favored in Cu, whereas Frank loops are favored in Al. The greater variability relative to the Ni results reflects the less extensive structural optimization performed for these comparative calculations.

*S3. Diffusive behavior of small disordered interstitial clusters*

The diffusion coefficients reported for the small disordered interstitial clusters $N$ = 1–5 were obtained from the linear-in-time regime of the cluster mean-squared displacement (MSD). Fig. S3 shows the MSD as a function of time at 900 K on logarithmic axes. For all cluster sizes examined, the MSD exhibits a log-log slope close to unity over the time interval analyzed, indicating diffusive behavior. The absence of significant deviations from linearity confirms that the simulated trajectories have reached the diffusive regime and that the diffusion coefficients reported in Fig. 5 of the main text are representative of long-time cluster transport.

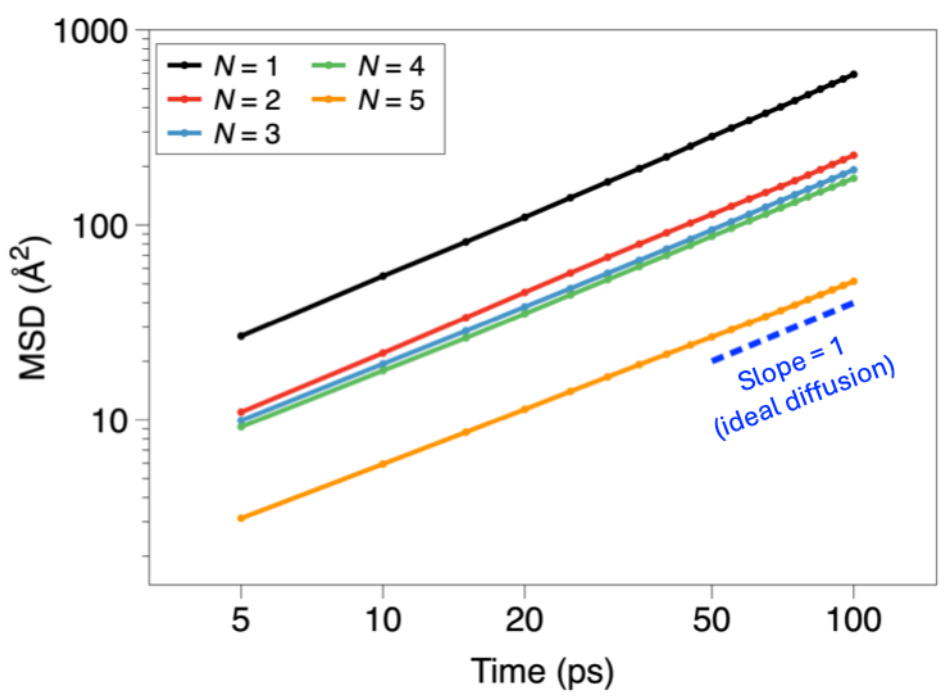

**Fig. S3.** Mean-squared displacement (MSD) of small disordered interstitial clusters in Ni at 900 K. Results are shown for cluster sizes $N$ = 1–5. The approximately linear behavior on the log-log plot corresponds to an MSD that grows proportionally with time, consistent with normal diffusive motion. The diffusion coefficients reported in Fig. 5 of the main text were obtained from this diffusive regime.

*S4. Diffusive behavior of compact, energetically stable perfect loops*

The diffusion coefficients reported in Fig. 7 of the main text were obtained from the center-of-mass (COM) mean-squared displacement (MSD) of compact, energetically stable perfect loops. Fig. S4 shows the evolution of the COM MSD for the two limiting cluster sizes investigated, $N = 16$ and $N = 400$, over the temperature range 400–1000 K. In both cases, the MSD exhibits a log-log slope close to unity over the time interval analyzed, confirming that the loops have reached the diffusive regime. The dashed line represents the ideal diffusive scaling, MSD $\propto t$. Similar behavior was observed for all intermediate cluster sizes. To maximize the statistical accuracy of the calculated diffusion coefficients, the MSD at each time interval was evaluated using all possible time origins within each trajectory and subsequently averaged over five independent simulations at each temperature. Diffusion coefficients were obtained from the one-dimensional Einstein relation using a displacement time interval of $\Delta t = 25$ ps.

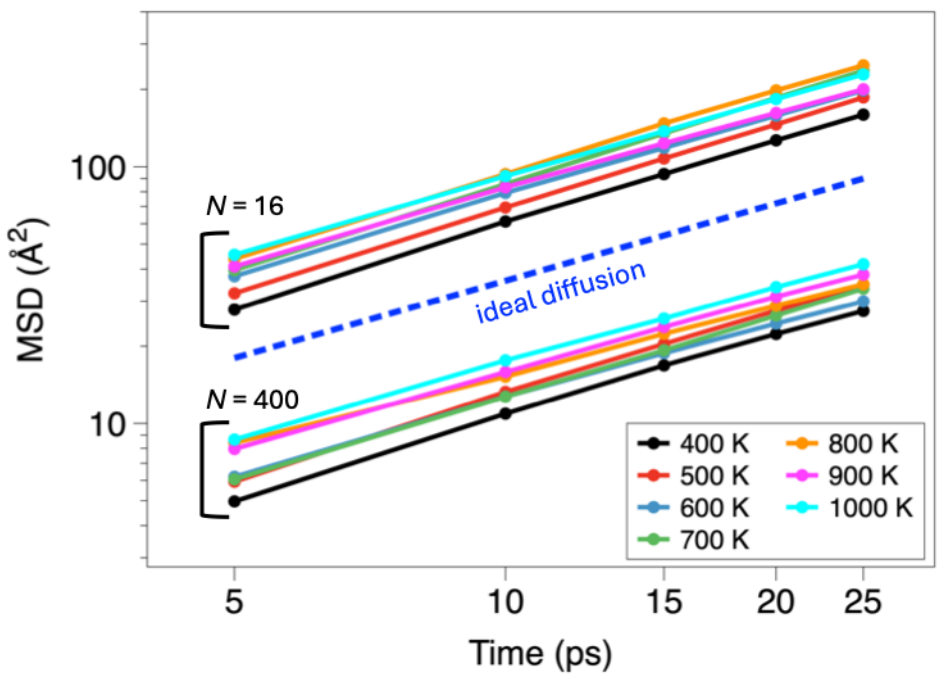

**Fig. S4.** Center-of-mass mean-squared displacement (MSD) of compact perfect loops for the smallest ($N = 16$) and largest ($N = 400$) cluster sizes investigated over the temperature range 400–1000 K. In both cases, the MSD increases approximately linearly with time, indicating one-dimensional diffusive motion. The dashed line illustrates the ideal diffusive behavior.

The diffusion coefficient is relatively insensitive to the displacement time interval used in the MSD analysis, and we used 25 ps for all the results in the main paper. Fig. S5(a) shows the diffusion coefficient of the largest loop ($N = 400$ at 1000 K), normalized by the value obtained at $\Delta t = 25$ ps, as a function of the displacement time interval. Over the range of intervals examined, the calculated diffusion coefficient varies by less than 20%, indicating that the extracted diffusivities are robust with respect to the choice of $\Delta t$. To assess the quality of the diffusive behavior, we also calculated $\mathrm{RMSE}_{s=1}$, defined as the root-mean-square deviation of the log-log MSD from a slope-one line. The selected interval of 25 ps lies in a regime with a small $\mathrm{RMSE}_{s=1}$, indicating close agreement with the linear scaling expected for diffusive motion. Fig. S5(b) shows the COM MSD of the $N = 400$ loop over 1 ns at 1000 K on a log-log scale. A slight deviation from slope-one behavior is observed at the longest times, which is consistent with the gradual increase in $\mathrm{RMSE}_{s=1}$ for larger displacement intervals shown in Fig. S5(a).

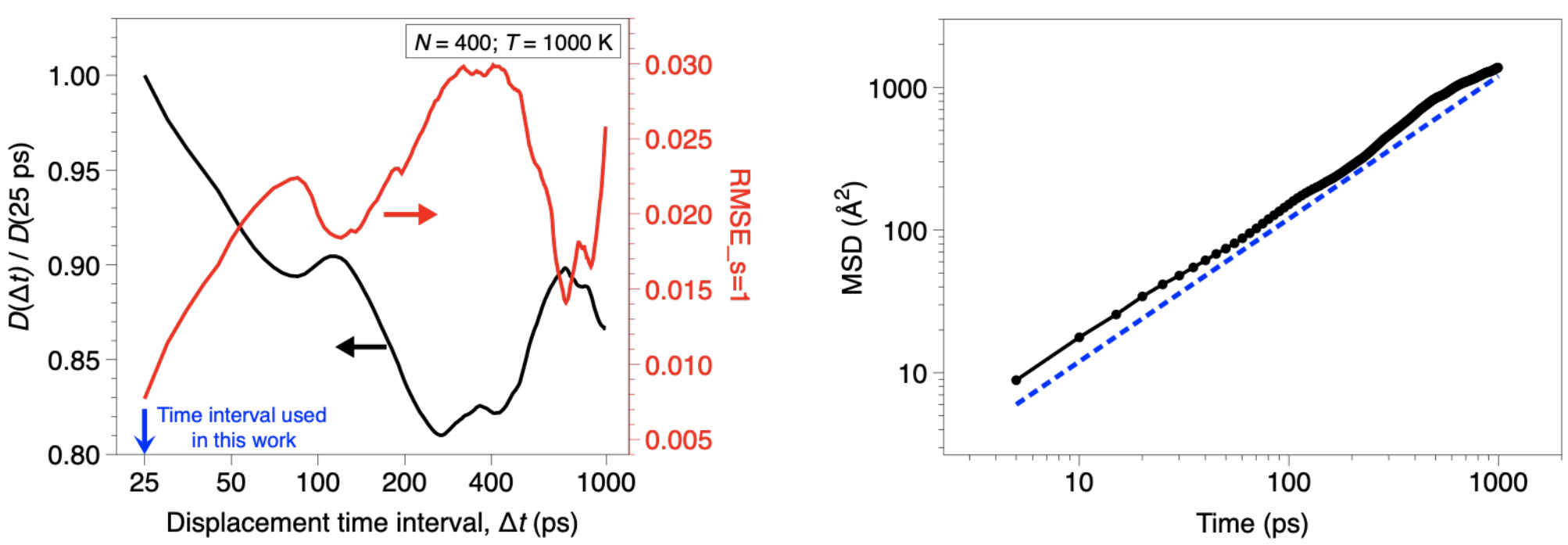

**Fig. S5.** Sensitivity of the COM diffusion coefficient to the displacement time interval used in the MSD analysis. (a) The black curve shows the effective diffusion coefficient obtained using different displacement intervals, normalized by the value at $\Delta t = 25$ ps. The red curve shows $\mathrm{RMSE}_{s=1}$, defined as the root-mean-square deviation of the log-log MSD data from a slope-one line. Smaller $\mathrm{RMSE}_{s=1}$ values indicate closer agreement with diffusive scaling. The 25 ps interval used in this work lies in a regime with small slope-one deviation and gives a diffusion coefficient consistent with values obtained over longer intervals. (b) COM MSD of the $N = 400$ perfect loop at 1000 K over 1 ns, illustrating increased deviation from the ideal diffusive behavior at long times.

*S5. Mass dependence of the diffusion coefficient*

To test the hypothesis that the dominant size dependence of the diffusion prefactor arises from the inverse-square-root mass dependence of the attempt frequency predicted by harmonic transition-state theory, additional MD simulations were performed for a perfect loop containing $N = 100$ SIAs. The atomic mass of Ni was artificially varied from one-quarter to four times its nominal value while keeping the interatomic potential unchanged. This procedure changes only the inertial mass of the atoms, allowing the mass dependence of the diffusion coefficient to be examined independently of the underlying energy landscape. Fig. S6 shows the resulting diffusion coefficients as a function of atomic mass. The data are well described by a power-law relation, $D \propto m^{-0.52\pm0.03}$, which is in excellent agreement with the harmonic prediction, $D \propto m^{-1/2}$. These results provide direct numerical support for the assumption that the attempt frequency scales inversely with the square root of the effective moving mass. Consequently, the observed decrease in the diffusion prefactor with increasing cluster size is consistent with the increase in the effective mass participating in each migration event.

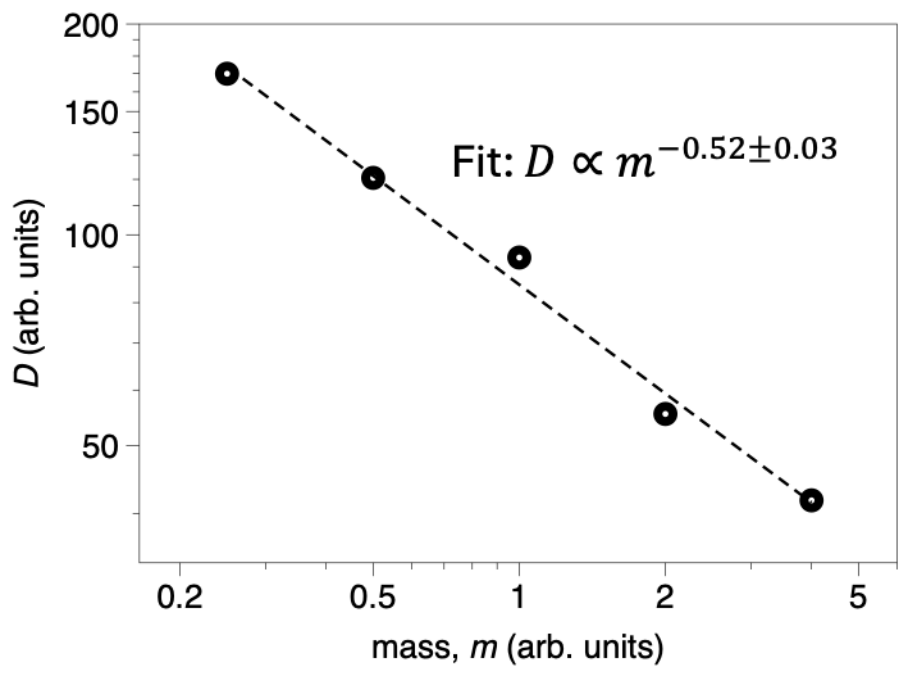

**Fig. S6.** Dependence of the diffusion coefficient of a perfect loop containing $N = 100$ SIAs on the atomic mass of Ni. The atomic mass was varied from one-quarter to four times its nominal value while keeping the interatomic potential unchanged. The dashed line is a power-law fit to the simulation results, yielding $D \propto m^{-0.52\pm0.03}$, in excellent agreement with the harmonic prediction $D \propto m^{-1/2}$.

*S6. Robustness of the participating-dumbbell scaling*

The analysis presented in the main text used a displacement interval of $\Delta t = 1$ ps and identified participating dumbbells as those with projected displacements exceeding the nearest-neighbor spacing ($r_c = 2.2$ Å). To assess the sensitivity of the extracted scaling exponent to these analysis parameters, we repeated the calculations using alternative displacement intervals and displacement thresholds. Fig. S7 summarizes two representative tests. In Fig. S7(a), the displacement interval was increased from 1 ps to 5 ps while retaining the original displacement threshold of $r_c = 2.2$ Å. This yields a slightly larger scaling exponent, $\langle N_{\mathrm{move}} \rangle \propto N^{0.89}$. In Fig. S7(b), the original displacement interval ($\Delta t = 1$ ps) was retained, but the displacement threshold was reduced from 2.2 Å to 1.0 Å, resulting in a scaling exponent of $\langle N_{\mathrm{move}} \rangle \propto N^{0.80}$. Although the fitted exponent varies somewhat with the specific analysis parameters, the scaling remains consistently sub-linear over the range examined, with exponents between approximately 0.80 and 0.89. These results demonstrate that the principal conclusion of the main text – that the number of participating dumbbells increases systematically with cluster size but grows more slowly than linearly – is robust with respect to reasonable variations in the analysis procedure.

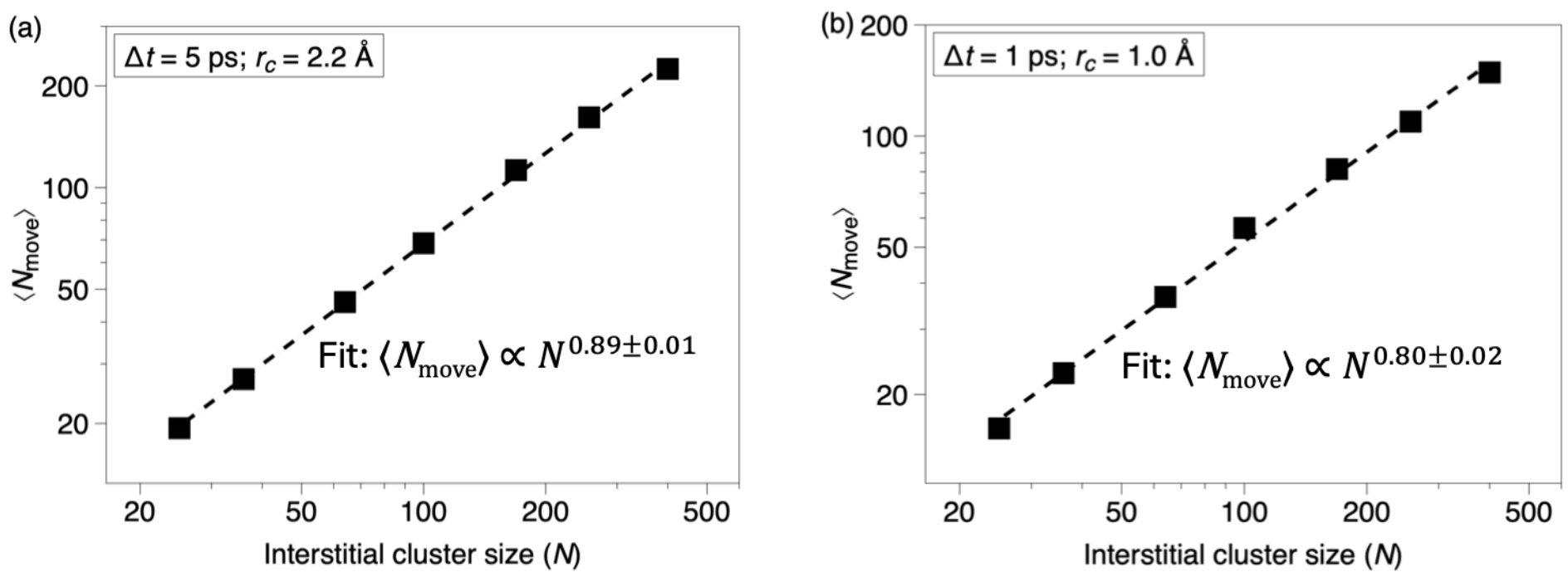

**Fig. S7.** Sensitivity of the scaling of the average number of participating dumbbells, $\langle N_{\mathrm{move}} \rangle$, to the analysis parameters. (a) Results obtained using a displacement interval of $\Delta t = 5$ ps and a displacement threshold of $r_c = 2.2$ Å. (b) Results obtained using $\Delta t = 1$ ps and a reduced displacement threshold of $r_c = 1.0$ Å. The fitted exponents vary from 0.80 to 0.89, demonstrating that the observed sub-linear scaling of $\langle N_{\mathrm{move}} \rangle$ with cluster size is robust to reasonable variations in the analysis parameters.

*S7. Estimation of the defect-loop diffusivity during pinning using confined Brownian motion*

During the pinning intervals identified in Fig. 10(b) of the manuscript, the defect loop does not remain at a fixed position. Instead, its measured position fluctuates over a finite spatial region along the migration direction. Within the model and assumptions described below, this intrawell motion can be used to estimate an effective diffusion coefficient for the pinned environment. Because the loop remains localized near a pinning site, treating these fluctuations as unrestricted one-dimensional diffusion is not appropriate. We therefore model the pinned loop as a Brownian particle undergoing one-dimensional diffusion within a finite reflecting interval (equivalently, a flat potential interval bounded by infinite walls). The purpose of this analysis is to determine the diffusion coefficient, *D*, that reproduces the experimentally measured mean absolute displacement, $\langle |\Delta x| \rangle$, between consecutive TEM frames separated by approximately 0.9 ms while accounting explicitly for the finite spatial extent of the pinning region.

We represent a pinning site as a one-dimensional interval of width $W$, with $0 \leq x \leq W$. Within the interval, the loop undergoes ordinary Brownian diffusion with diffusion coefficient *D*, and the boundaries at x = 0 and x = $W$ are reflecting. Let $p(x, t \mid x_0)$ denote the conditional probability density for finding the particle at x after elapsed time $\Delta t$, given that it started at $x_0$. This density satisfies the one-dimensional diffusion equation:

SUPPORTING INFORMATION

$$\frac{\partial p(x,t|x_0)}{\partial t} = D\frac{\partial^2 p(x,t|x_0)}{\partial x^2}.$$

Reflecting boundaries correspond to zero probability flux at the two walls:

$$\partial p/\partial x\,|_{x=0} = 0, \qquad \partial p/\partial x\,|_{x=W} = 0,$$

Physically, these conditions mean that the particle cannot escape the local pinning region. When Brownian motion carries it toward either boundary, it is reflected back into the interval. In the coordinate convention used here, with the interval extending from 0 to $W$, the propagator for Brownian diffusion in a one-dimensional box with reflecting boundaries is given by [6]:

$$p(x, \Delta t \mid x_0) = 1/W + (2/W)\sum_{n=1}^{\infty} \cos(n\pi x/W)\cos(n\pi x_0/W)\,\exp[-n^2\pi^2 D\Delta t/W^2]. \quad \text{(S1)}$$

The first term, $1/W$, is the equilibrium probability density. At sufficiently long times, all exponential terms decay to zero, so the position becomes uniformly distributed throughout the interval:

$$p(x, \Delta t \to \infty \mid x_0) = 1/W.$$

This long-time behavior has a simple physical interpretation: after sufficient time, the particle has completely lost memory of where it started and is equally likely to be found anywhere inside the reflecting interval.

The experimentally accessible quantities used here are the mean absolute displacement between consecutive TEM frames, $\langle|\Delta x|\rangle$, and the confinement width, W. For consecutive frames i and i + 1, $\Delta x_i = x_{i+1} - x_i$, and the measured mean absolute displacement is the mean of the observed $|\Delta x_i|$ values. The absolute value is used because the loop can fluctuate in either direction within the pinning region. For the three pinning intervals in Fig. 10(b), analysis of the raw trajectory gave mean absolute frame-to-frame displacements of approximately 1.10, 1.57, and 1.37 nm for intervals 1, 2, and 3, respectively. Pooling the successive displacements from all three intervals gives $\langle|\Delta x|\rangle \approx 1.44$ nm. The trajectories also show that the loop remains localized within a region several nanometers wide along the principal migration direction. Because the apparent range of a finite trajectory can be strongly influenced by extreme positions and slow motion of the center of the pinning region, the observed maximum-minus-minimum range should not be interpreted as an exact hard-wall separation. We therefore use $W \approx 5$ nm as a characteristic confinement width, consistent with the spatial extent of the observed fluctuations.

For unrestricted one-dimensional Brownian motion, the displacement over time $\Delta t$ is Gaussian with variance $2D\Delta t$, and its mean absolute value is $\langle|\Delta x|\rangle = \sqrt{4D\Delta t/\pi}$. In the present case, however, the loop is confined to a region of width $W$, so the frame-to-frame displacement cannot grow indefinitely as $D$ increases. The limiting cases provide useful checks. As $D \to 0$, $\langle|\Delta x|\rangle \to 0$. As $D \to \infty$, consecutive positions become independent uniform random variables on $[0, W]$, whose mean absolute separation is $W/3$. Thus, the decorrelated limit is $\langle|\Delta x|\rangle = W/3 = 1.67$ nm for $W = 5$ nm. The measured value, 1.44 nm, lies below this limit, so consecutive positions are not completely decorrelated and retain information about a finite diffusion coefficient.

The expected frame-to-frame displacement is obtained by averaging $|x - x_0|$ over all initial and final positions, weighted by their probabilities:

$$\langle|\Delta x|\rangle = \int_0^W dx_0\, \rho(x_0) \int_0^W dx\, |x - x_0|\, p(x,\Delta t| x_0). \quad \text{(S2)}$$

Here $\rho(x_0)$ is the equilibrium probability density of the initial position. For a flat reflecting interval, the equilibrium density is uniform, $\rho(x_0) = 1/W$. Equation (S2) therefore connects the measured mean absolute displacement directly to the propagator. Equation (S2) can be evaluated analytically. Defining $u = x/W$, $u_0 = x_0/W$, $\alpha = D\Delta t/W^2$, and the dimensionless propagator $q(u, \alpha \mid u_0) = W p(x, \Delta t \mid x_0)$, equation (S1) becomes:

$$q(u, \alpha \mid u_0) = 1 + 2\sum_{n=1}^{\infty} \cos(n\pi u)\cos(n\pi u_0)\exp(-n^2\pi^2\alpha).$$

Substituting this expansion into Eq. (S2), the constant term gives $W/3$. Each nonconstant eigenmode can be integrated using

$$\int_0^1 du_0 \int_0^1 du\, |u-u_0| \cos(n\pi u)\cos(n\pi u_0) = -1/(n^2\pi^2).$$

The analytical relation between the measured displacement and the dimensionless diffusion parameter is therefore

$$f(\alpha) \equiv \frac{\langle|\Delta x|\rangle}{W} = \frac{1}{3} - \frac{2}{\pi^2}\sum_{n=1}^{\infty} \exp(-n^2\pi^2\alpha)/n^2. \quad \text{(S3)}$$

This expression has the required limits: $f(0) = 0$ because $\sum_{n=1}^{\infty} 1/n^2 = \pi^2/6$, and $f(\alpha) \to 1/3$ as $\alpha \to \infty$. Moreover, $f'(\alpha) = 2\sum_{n=1}^{\infty} \exp(-n^2\pi^2\alpha) > 0$, so every measured ratio below 1/3 corresponds to a unique $\alpha$ within this model. For the experiment, $\langle|\Delta x|\rangle/W = 1.44/5.0 = 0.288$. Solving Eq. (S3) gives $\alpha \approx 0.152$. As an independent numerical check, Brownian dynamics with $3 \times 10^6$ pairs – uniform $u_0$ between 0 and 1, a Gaussian free displacement $\sqrt{2\alpha}\,\xi$ with $\xi \sim N(0, 1)$, and repeated mirror reflections – gives 0.28799 ± 0.00013 (standard error), in agreement with the analytical value 0.28800. Substitution in the values for $\alpha$, $W = 5$ nm, and $\Delta t = 0.9$ ms gives $D \approx 4.2 \times 10^{-15}$ m$^2$ s$^{-1}$.

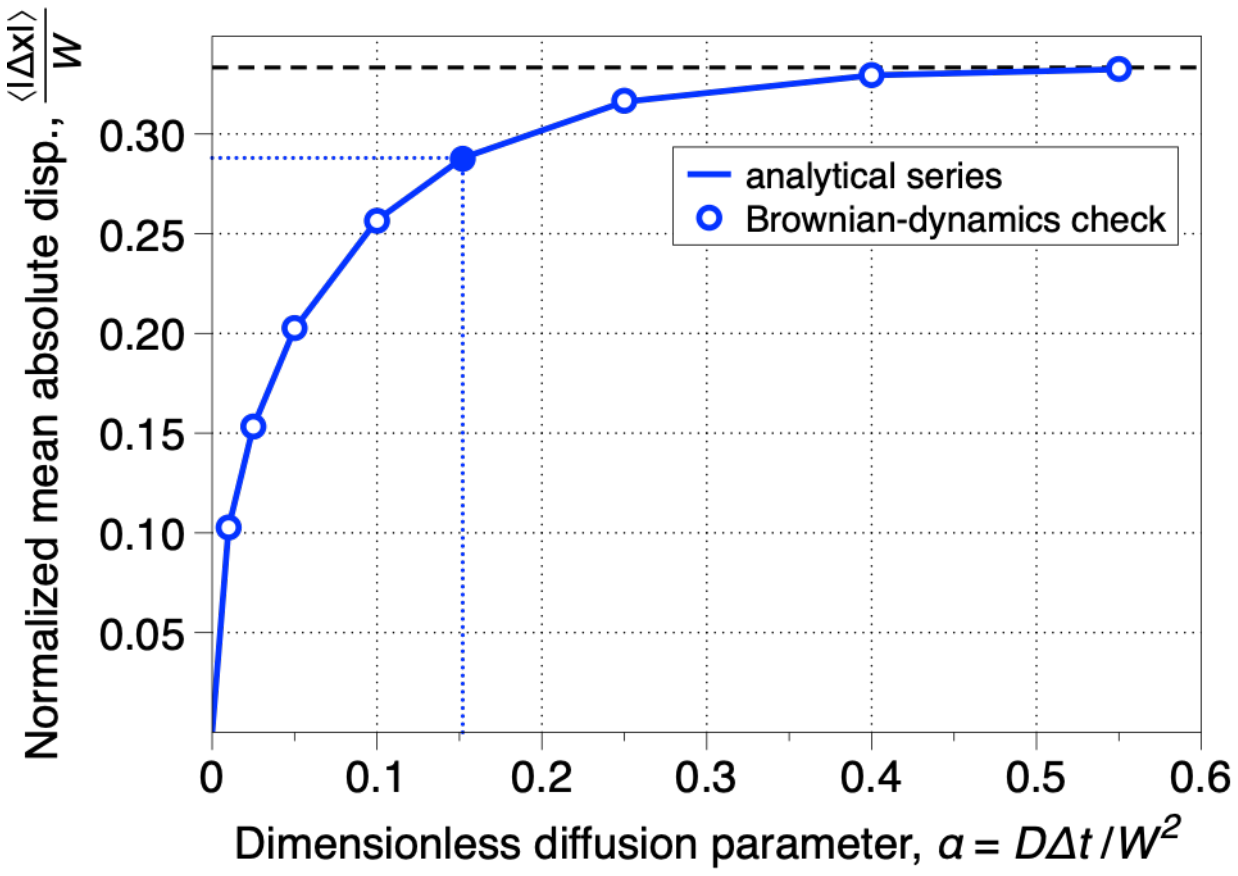

Fig. S8. Normalized mean absolute displacement, $\langle|\Delta x|\rangle/W$, as a function of $\alpha = D\Delta t/W^2$ for one-dimensional diffusion in a reflecting interval. The solid curve is the analytical series in Eq. (S3), and the open circles are Brownian-dynamics checks. The dashed horizontal line marks the decorrelated limit of 1/3. The experimental ratio, 0.288, gives $\alpha = 0.151999 \approx 0.152$.

Thus, within the one-dimensional reflecting-well approximation, the pinning-segment fluctuations correspond to an effective intrawell diffusivity of approximately $4.2 \times 10^{-15}$ m$^2$ s$^{-1}$. This estimate is conditional on a fixed confinement width, equilibrium sampling within a flat well, and negligible localization error and motion blur; departures from these assumptions would change the inferred value.

## References

[1] P.W. Ma, S.L. Dudarev, Nonuniversal structure of point defects in face-centered cubic metals, Phys. Rev. Mater. 5 (2021) 013601. https://doi.org/10.1103/PHYSREVMATERIALS.5.013601/FIGURES/27/MEDIUM.

[2] S. Bukkuru, U. Bhardwaj, K.S. Rao, A.D.P. Rao, M. Warrier, M.C. Valsakumar, Kinetics of self-interstitial migration in bcc and fcc transition metals, Mater. Res. Express 5 (2018) 035513. https://doi.org/10.1088/2053-1591/AAB418.

[3] Y.N. Osetsky, A. V. Barashev, Y. Zhang, On the mobility of defect clusters and their effect on microstructure evolution in fcc Ni under irradiation, Materialia (Oxf). 4 (2018) 139–146. https://doi.org/10.1016/J.MTLA.2018.09.028.

[4] M.I. Mendelev, M.J. Kramer, C.A. Becker, M. Asta, Analysis of semi-empirical interatomic potentials appropriate for simulation of crystalline and liquid Al and Cu, Philosophical Magazine 88 (2008) 1723–1750. https://doi.org/10.1080/14786430802206482.

[5] A.M. Goryaeva, C. Domain, A. Chartier, A. Dézaphie, T.D. Swinburne, K. Ma, M. Loyer-Prost, J. Creuze, M.C. Marinica, Compact A15 Frank-Kasper nano-phases at the origin of dislocation loops in face-centred cubic metals, Nature Communications 2023 14:1 14 (2023) 3003-. https://doi.org/10.1038/s41467-023-38729-6.

[6] K.I. Mortensen, H. Flyvbjerg, J.N. Pedersen, Confined Brownian Motion Tracked With Motion Blur: Estimating Diffusion Coefficient and Size of Confining Space, Front. Phys. 8 (2021) 583202. https://doi.org/10.3389/FPHY.2020.583202/TEXT.